\documentclass[twocolumn,openany,oneside]{aastex63}
\usepackage{graphicx,amsmath,amssymb,color,appendix,soul}
\usepackage{footnote,tabularx}

\newcommand{\Teff}{\mbox{$T_{\mathrm{eff}}$}}
\newcommand{\logg}{\mbox{$\log g$}}
\newcommand{\feh}{\mbox{$\mathrm{[Fe/H]}$}}

\newcommand{\kms}{\mbox{$\mathrm{km\,s^{-1}}$}}
\newcommand{\Ion}[2]{#1{\,\scriptsize #2}}

\defcitealias{Parsons2016MNRAS.463.2125P}{Paper~I}
\defcitealias{Rebassa2017MNRAS.472.4193R}{Paper~II}
\defcitealias{Lagos2020MNRAS.494..915L}{Paper~III}

\begin{document}

\title{The white dwarf binary pathways survey\\ V. The \emph{Gaia} white dwarf plus AFGK binary sample and the identification of 23 close binaries}

\author[0000-0003-3243-464X]{J.-J.~Ren}
\affil{CAS Key Laboratory of Space Astronomy and Technology, National Astronomical Observatories, Chinese Academy of Sciences, Beijing 100101, China}
\email{jjren@nao.cas.cn}

\author[0000-0002-9090-9191]{R.~Raddi}
\affil{Departament de F\'isica, Universitat Polit\`ecnica de Catalunya, c/Esteve Terrades 5, E-08860 Castelldefels, Spain}

\author[0000-0002-6153-7173]{A.~Rebassa-Mansergas}
\affil{Departament de F\'isica, Universitat Polit\`ecnica de Catalunya, c/Esteve Terrades 5, E-08860 Castelldefels, Spain}
\affil{Institute for Space Studies of Catalonia, c/Gran Capit\`a 2--4, Edif. Nexus 201, 08034 Barcelona, Spain}

\author{M.~S.~Hernandez}
\affil{Instituto de F\'\i  sica  y  Astronom\'\i a,  Universidad  de  Valpara\'\i so, Avenida Gran Breta\~na 1111, Valpara\'\i so, Chile}

\author[0000-0002-2695-2654]{S.~G.~Parsons}
\affil{Department of Physics and Astronomy, University of Sheffield, Sheffield S3 7RH, UK}

\author{P.~Irawati}
\affil{National Astronomical Research Institute of Thailand, Sirindhorn AstroPark, Donkaew, Mae Rim, Chiang Mai 50180, Thailand}

\author{P.~Rittipruk}
\affil{National Astronomical Research Institute of Thailand, Sirindhorn AstroPark, Donkaew, Mae Rim, Chiang Mai 50180, Thailand}

\author[0000-0003-3903-8009]{M.~R.~Schreiber}
\affil{Departamento de F{\'i}sica, Universidad T\'ecnica Federico Santa Mar\'ia, Av. Espa\~{n}a 1680, Valpara{\'i}so, Chile}
\affil{Millennium Nucleus for Planet formation, NPF,  Valpara{\'i}so, Chile}

\author[0000-0002-2761-3005]{B.~T.~G\"ansicke}
\affil{Department of Physics, University of Warwick, Coventry CV4 7AL, UK}
\affil{Centre for Exoplanets and Habitability, University of Warwick, Coventry CV4 7AL, UK}

\author{S.~Torres}
\affil{Departament de F\'isica, Universitat Polit\`ecnica de Catalunya, c/Esteve Terrades 5, E-08860 Castelldefels, Spain}
\affil{Institute for Space Studies of Catalonia, c/Gran Capit\`a 2--4, Edif. Nexus 201, 08034 Barcelona, Spain}

\author{H.-J.~Wang}
\affil{CAS Key Laboratory of Optical Astronomy, National Astronomical Observatories, Chinese Academy of Sciences, Beijing 100101, China}
\affil{University of Chinese Academy of Sciences, Beijing 100049, China}

\author{J.-B.~Zhang}
\affil{CAS Key Laboratory of Optical Astronomy, National Astronomical Observatories, Chinese Academy of Sciences, Beijing 100101, China}

\author{Y.~Zhao}
\affil{CAS Key Laboratory of Optical Astronomy, National Astronomical Observatories, Chinese Academy of Sciences, Beijing 100101, China}

\author{Y.-T.~Zhou}
\affil{Department of Astronomy, Peking University, Beijing 100871, China}
\affil{Kavli Institute for Astronomy and Astrophysics, Peking University, Beijing 100871, China}

\author{Z.-W.~Han}
\affil{Key Laboratory for the Structure and Evolution of Celestial Objects, Yunnan observatories, Chinese Academy of Sciences, PO Box 110, Kunming, 650011, Yunnan Province, China}

\author{B.~Wang}
\affil{Key Laboratory for the Structure and Evolution of Celestial Objects, Yunnan observatories, Chinese Academy of Sciences, PO Box 110, Kunming, 650011, Yunnan Province, China}

\author[0000-0002-1802-6917]{C.~Liu}
\affil{CAS Key Laboratory of Space Astronomy and Technology, National Astronomical Observatories, Chinese Academy of Sciences, Beijing 100101, China}
\affil{University of Chinese Academy of Sciences, Beijing 100049, China}

\author{X.-W.~Liu}
\affil{South-Western Institute for Astronomy Research, Yunnan University, Kunming, Yunnan, 650091, China}

\author{Y.~Wang}
\affil{CAS Key Laboratory of Optical Astronomy, National Astronomical Observatories, Chinese Academy of Sciences, Beijing 100101, China}

\author{J.~Zheng}
\affil{CAS Key Laboratory of Optical Astronomy, National Astronomical Observatories, Chinese Academy of Sciences, Beijing 100101, China}

\author{J.-F.~Wang}
\affil{CAS Key Laboratory of Optical Astronomy, National Astronomical Observatories, Chinese Academy of Sciences, Beijing 100101, China}
\affil{University of Chinese Academy of Sciences, Beijing 100049, China}

\author{F.~Zhao}
\affil{CAS Key Laboratory of Optical Astronomy, National Astronomical Observatories, Chinese Academy of Sciences, Beijing 100101, China}

\author{K.-M.~Cui}
\affil{CAS Key Laboratory of Optical Astronomy, National Astronomical Observatories, Chinese Academy of Sciences, Beijing 100101, China}
\affil{University of Chinese Academy of Sciences, Beijing 100049, China}

\author{J.-R.~Shi}
\affil{CAS Key Laboratory of Optical Astronomy, National Astronomical Observatories, Chinese Academy of Sciences, Beijing 100101, China}
\affil{University of Chinese Academy of Sciences, Beijing 100049, China}

\author[0000-0003-3347-7596]{H.~Tian}
\affil{CAS Key Laboratory of Space Astronomy and Technology, National Astronomical Observatories, Chinese Academy of Sciences, Beijing 100101, China}

\begin{abstract}
Close white dwarf binaries consisting of a white dwarf and an A, F, G or K type main sequence star, henceforth close WD+AFGK binaries, are ideal systems to understand the nature of type Ia supernovae progenitors and to test binary evolution models. In this work we identify 775 WD+AFGK candidates from TGAS (The \emph{Tycho}-\emph{Gaia} Astrometric Solution) and \emph{Gaia} Data Release 2 (DR2), a well-defined sample of stars with available parallaxes, and we measure radial velocities (RVs) for 275 of them with the aim of identifying close binaries. The RVs have been measured from high resolution spectra obtained at the Xinglong 2.16\,m Telescope and the San Pedro M\'artir 2.12\,m Telescope and/or from available LAMOST DR6 (low-resolution) and RAVE DR5 (medium-resolution) spectra. We identify 23 WD+AFGK systems displaying more than 3$\sigma$ RV variation among 151 systems for which the measured values are obtained from different nights. Our WD+AFGK binary sample contains both AFGK dwarfs and giants, with a giant fraction $\sim$\,43\%. The close binary fractions we determine for the WD+AFGK dwarf and giant samples are $\simeq$\,24\,\% and $\simeq$15\,\%, respectively. We also determine the stellar parameters (i.e. effective temperature, surface gravity, metallicity, mass and radius) of the AFGK companions with available high resolution spectra. The stellar parameter distributions of the AFGK companions that are members of close and wide binary candidates do not show statistically significant differences.
\end{abstract}


\section{Introduction}

A large fraction of stars are born in binary systems \citep{Duchene2013ARA&A..51..269D}. Thus, the study of binary evolution represents an important part in our understanding of stellar evolution. The majority ($\simeq$ 75\,\%) of medium/low-mass main sequence binaries have relatively large orbital separations, evolving as if they were single stars and never interacting \citep{deKool1992A&A...261..188D, Willems2004A&A...419.1057W}. The remaining $\simeq$\,25\,\% are believed to undergo dynamically unstable mass transfer episodes, when the more massive star evolves into a red giant or asymptotic giant, which generally results in a common envelope (CE) phase \citep{Paczynski1976IAUS...73...75P, Iben1993PASP..105.1373I, Webbink2008ASSL..352..233W}. After the ejection of the CE \citep{Passy2012ApJ...744...52P, Ricker2012ApJ...746...74R}, a close post-CE binary (PCEB) is formed, which contains a compact object, usually a white dwarf (WD, that is the former core of the giant star) and a main-sequence (MS) star companion.

PCEBs evolve to shorter orbital periods through angular momentum loss driven by magnetic braking and/or gravitational wave emission. Depending upon the stellar masses and orbital separations, PCEBs may undergo a second CE stage, presumably leading to double-degenerate binaries \citep{Nelemans2001A&A...365..491N, Rebassa2017MNRAS.466.1575R, Breedt2017MNRAS.468.2910B, Kilic2017MNRAS.471.4218K}; or enter a semi-detached state and become cataclysmic variables \citep[CVs;][]{Gansicke2009MNRAS.397.2170G, Pala2017MNRAS.466.2855P} or super-soft X-ray sources \citep[SSSs;][]{Kalomeni2016ApJ...833...83K, Parsons2015MNRAS.452.1754P}. Double degenerate binaries, CVs (most probably the recurrent novae) and SSSs are of high interest, since they are considered to be the possible progenitors of Type Ia supernova \citep[SN Ia;][]{Langer2000A&A...362.1046L}, see \citet{Parthasarathy2007NewAR..51..524P} and \citet{Wang2012NewAR..56..122W} for a review of the various types of promising observed SN Ia progenitors. 

Although it is well established that SN Ia are related to the thermonuclear ignition of a C/O core WD, there is not yet a general consensus on the pathways leading to the explosion \citep{Han2004MNRAS.350.1301H, Wang2010MNRAS.401.2729W, Wang2018RAA....18...49W}. This may limit the use of SN Ia as cosmological probes \citep{Riess1998AJ....116.1009R, Perlmutter1999ApJ...517..565P}. Up to five different scenarios have been proposed leading to the explosion of SN Ia: the single- and double-degenerate scenario, the core degenerate scenario, the double-detonation and the WD+WD collision scenario \citep[see][and references therein]{Soker2018SCPMA..61d9502S}. It is likely that all these channels contribute to the observed SN Ia population, but no current population synthesis model can reproduce both the observed SN Ia rates and delay time distributions \citep{Maoz2012PASA...29..447M, Maoz2012MNRAS.426.3282M, Wang2012NewAR..56..122W}.

Depending on the mass ratio and orbital period, PCEBs with more massive (A, F, G and early-K spectral type) MS secondary stars may either evolve through a second CE and become double-degenerates \citep{Wang2012NewAR..56..122W}, or some may begin mass transfer (depends on mass retention) to explode as near/sub-Chandrasekhar SN Ia \citep{Whelan1973ApJ...186.1007W, Flors2020MNRAS.491.2902F}. Therefore, WD+AFGK PCEBs hold the potential to statistically test which progenitor channel is more efficient for producing SN Ia, i.e. observations of detached close WD+AFGK binaries offer us the potential to simultaneously sample the progenitors of many SN Ia channels.

We have initiated a large-scale observational project dedicated to (1) identify a large sample of WD+AFGK candidates and to (2) search for systems displaying significant radial velocity (RV) variations, i.e. likely close WD+AFGK PCEBs, for measuring their orbital periods. Our goals are both predicting the future of these close binary systems to test possible SN Ia progenitor channels and constraining their past evolution to test the dependence of CE efficiency with the secondary star mass. To identify WD+AFGK binaries we have previously mined the Radial Velocity Experiment (RAVE) \citep{Kordopatis2013AJ....146..134K} and LAMOST surveys. The detailed identification methodology was presented in \citet[Paper I;][]{Parsons2016MNRAS.463.2125P}, which mainly uses a \Teff\ (effective temperature of the MS companion) vs. FUV$-$NUV (the {\em GALEX} far-UV minus near-UV colour) selection criteria. Until now, 430 WD+AFGK binaries have been identified from  RAVE data release (DR) 4 \citepalias{Parsons2016MNRAS.463.2125P}, and 1549 from LAMOST DR4 \citep[Paper II;][]{Rebassa2017MNRAS.472.4193R}. Follow-up spectroscopic observations have allowed the identification of 24 close LAMOST WD+AFGK systems displaying RV variation above the 3-$\sigma$ level \citepalias{Rebassa2017MNRAS.472.4193R}. The first hierarchical triple system in this White Dwarf Binary Pathways Survey, and the limited and acceptable fraction of contamination of WD+AFGK sample from hierarchical triple systems containing a WD is presented in \citet[Paper III;][]{Lagos2020MNRAS.494..915L}. The final goal of this project is to present a large number of WD+AFGK binaries with measured orbital periods, see Hernandez et al. (Paper IV; 2020, submitted) for the first results.

In this paper, we expand our search of WD+AFGK binaries by harvesting from the {\em Tycho}-\emph{Gaia} (hereafter TGAS) catalogue of stars with measured parallaxes in the first $Gaia$ data release \citep[DR1;][]{Lindegren2016A&A...595A...4L, Michalik2015A&A...574A.115M}. This sample, which we later complemented with the observations from the \emph{Gaia} Data Release 2 \citep[DR2;][]{Gaia2018A&A...616A...1G}, offers a great potential to study a statistically significant number of systems and a step towards a volume complete sample that can be built with the forthcoming third \emph{Gaia} data release. We then describe our high-resolution follow-up spectroscopic observations performed at the Xinglong 2.16\,m and San Pedro M\'artir (SPM) 2.12\,m telescopes. We determine their RVs and present the identified close binaries. We also measure the stellar parameters of the AFGK companions from our spectra and perform a statistical analysis of the properties of close and wide binary candidates. 

\section{The WD+AFGK sample}

\subsection{Preliminary selection from the Tycho-Gaia astrometric solution catalogue}
The first release of the $Gaia$ mission was presented in September 2016 \citep{Gaia2016A&A...595A...1G}, which marked the beginning of a new era in astrometry. The combination of the {\em Tycho}-2 catalogue \citep{Hog2000A&A...355L..27H} with {\em Gaia} DR1 \citep{Gaia2016A&A...595A...1G} led to the {\em Tycho}-{\em Gaia} astrometric solution \citep[TGAS;][]{Michalik2015A&A...574A.115M, Gaia2016A&A...595A...2G, Lindegren2016A&A...595A...4L} catalogue, which provides positions, photometry in the broad visual $G$ band, proper motions and parallaxes with typical accuracy (uncertainties in parallaxes typically under 1\,mas) of Hipparcos level or better for about two million stars up to $\sim$\,11.5\,mag. 

With the aim of identifying WD+AFGK binaries within TGAS, we cross-matched a sub-set of sources, limited to a parallax precision of $\sigma_\varpi/\varpi \leq 0.15$, with the  latest \emph{GALEX} release  \citep[$FUV \leq 19$\,mag;][]{Bianchi2017ApJS..230...24B}. The resulting $\approx$ 13\,000 sources were complemented with optical to infrared photometry from the AAVSO Photometric All-Sky Survey \citep[APASS DR9;][]{Henden2015AAS...22533616H}, Two-Micron All-Sky Survey \citep[2MASS;][]{Skrutskie2006AJ....131.1163S}, and the \emph{Wide-field Infrared Survey Explorer} \citep[\emph{WISE;}][]{Wright2010AJ....140.1868W}. This sample was later supplemented, and re-analyzed, with the addition of {\em Gaia} DR2 photometry and parallaxes (see Section 2.2).

Using a grid of synthetic spectra from the PHOENIX library \citep{Husser2013A&A...553A...6H}\footnote{For the synthetic photometry we used the $Tycho$ band-passes and zero-points, and 2MASS zero-points determined by \citet{Maiz-apellaniz2006AJ....131.1184M}, APASS zero-points by \citet{Munari2014AJ....148...81M}, and {\em WISE} zero-points as prescribed in the survey paper.}, we estimated effective temperatures (\Teff), stellar radii ($R$), and interstellar extinction ($A_V$), via $\chi^2$ minimization of the observed and modelled spectral energy distributions (SED). Surface gravity $(\log{g}$) and metallicity ([Fe/H)] are also fitted, for the benefit of reaching the best fit, although the former is typically poorly determined and the latter is a-priori constrained to vary around $0.0 \pm0.5$\,dex; hence, they are discarded as reliable estimates. For the SED fitting, we also used the TGAS parallaxes and the total-column of dust \citep{Schlegel1998ApJ...500..525S} as external constraints, and we adopted the $R_V = 3.1$ wavelength-dependent reddening law of \citet{Fitzpatrick1999PASP..111...63F}. Initially we undertook a \emph{brute force} approach, evaluating the $\chi^2$ at each point of a tailored grid of models, followed by a Nelder-Mead method to identify the best-fit. Both methods are implemented in the \textsc{python}'s \textsc{scipy} \citep{scipy2020} and \textsc{lmfit} packages. The fitting procedure excluded the \emph{GALEX} $FUV$ and $NUV$ bands, where the optically brighter AFGK stars are assumed to contribute less than their WD companions.

\begin{figure}
   \centering
   \includegraphics[width=0.475\textwidth]{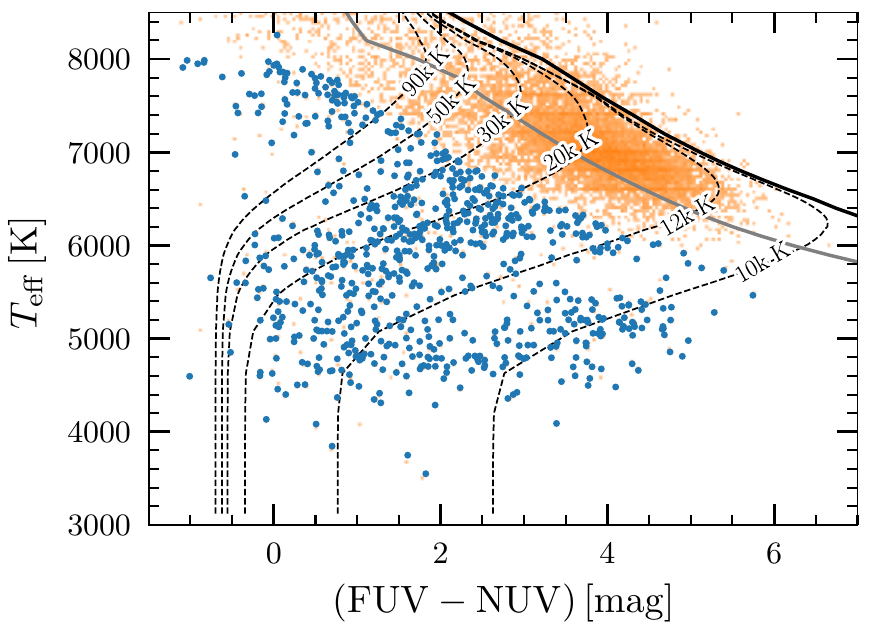}
   \caption{UV colours and photometric temperatures of $\approx 13\,000$ TGAS stars (orange cloud). Note that we only show 775 TGAS-selected candidates (blue symbols), resulting from our improved classification with Gaia DR2 and the cross-match with SIMBAD. The black and grey, solid curves represent the intrinsic colours of single stars with $\log{g} = 4.5,\,3.5$ and ${\rm [Fe/H]} = 0,\,-1$, respectively, determined for the PHOENIX grid of synthetic models \citep{Husser2013A&A...553A...6H}. The dashed black curves represent the intrinsic colours of unresolved WD+AFGK binaries for a range of WD temperatures (shown in figure) determined from $\log{g} = 8$ synthetic models \citep{Koester2010MmSAI..81..921K}.}
   \label{fig:colorcut}
\end{figure}

Thus, adopting a similar, although more relaxed selection criterion as defined in \citetalias{Parsons2016MNRAS.463.2125P}, we identified as possible WD+AFGK candidates 985 stars with $\Teff \leq 8000$\,K and 1.5\,mag bluer than the intrinsic FUV\,$-$\,NUV colours of PHOENIX models with $\log{g} = 3.5$ and \feh\,$= -1$. This cut enables us to remove typically metal-poor, non-binary systems that constitute the majority of the TGAS sample (see Figure\,\ref{fig:colorcut}).

\begin{figure}
   \centering
   \includegraphics[width=0.475\textwidth]{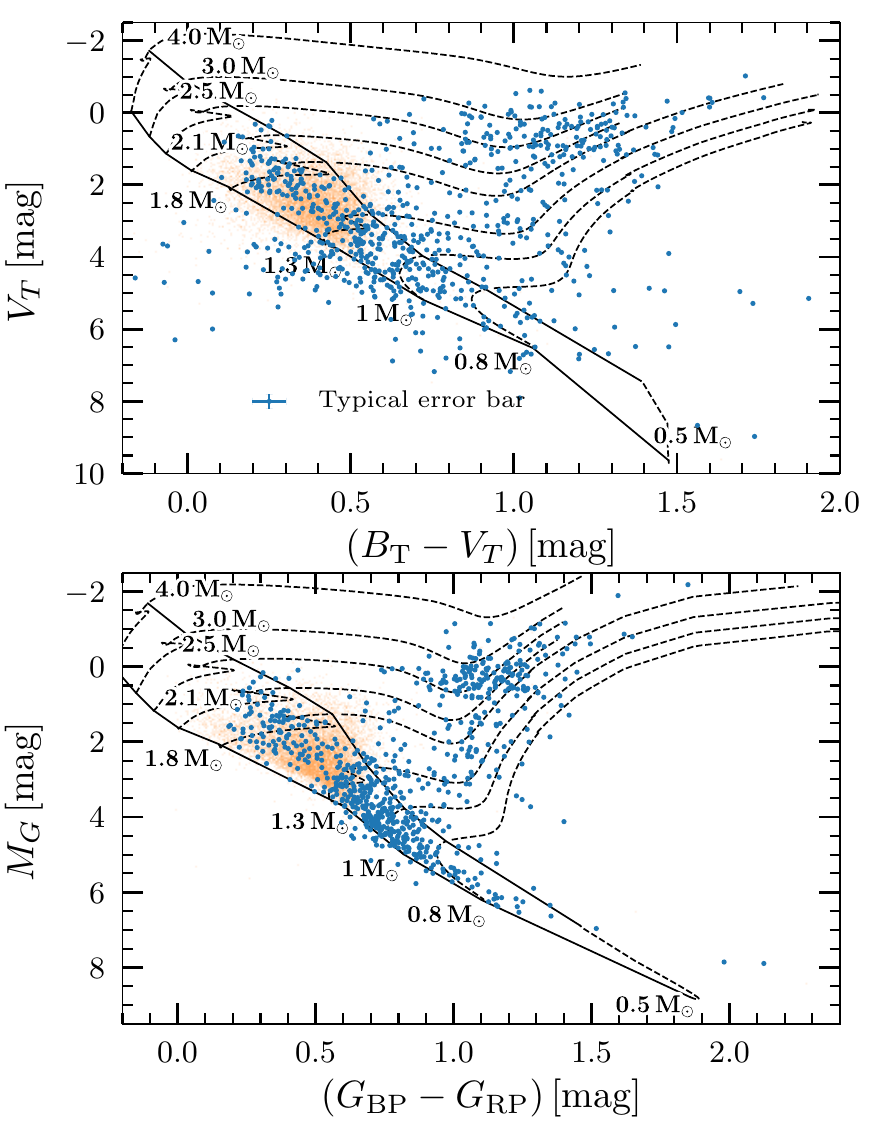} 
   \caption{The HR diagrams for the $\approx$ 13\,000 cross-matched TGAS-\emph{GALEX} stars (orange dots), which use the TGAS and \emph{Gaia} DR2 data (top and bottom panels, respectively). Symbols and colours as in Figure\,\ref{fig:colorcut}. The overlaid dashed curves are the MESA/MIST evolutionary tracks, while the black solid curves represent the zero-age and the terminal-age main sequences \citep[ZAMS and TAMS, respectively;][]{Choi2016ApJ...823..102C}. In Table\,\ref{tab:tgas-wdfgk-sample} we label as dwarfs/giants all stars below/above the TAMS track.}
   \label{fig:hr_tgas_dr2}
\end{figure}

\begin{figure}
   \centering
   \includegraphics[width=0.475\textwidth]{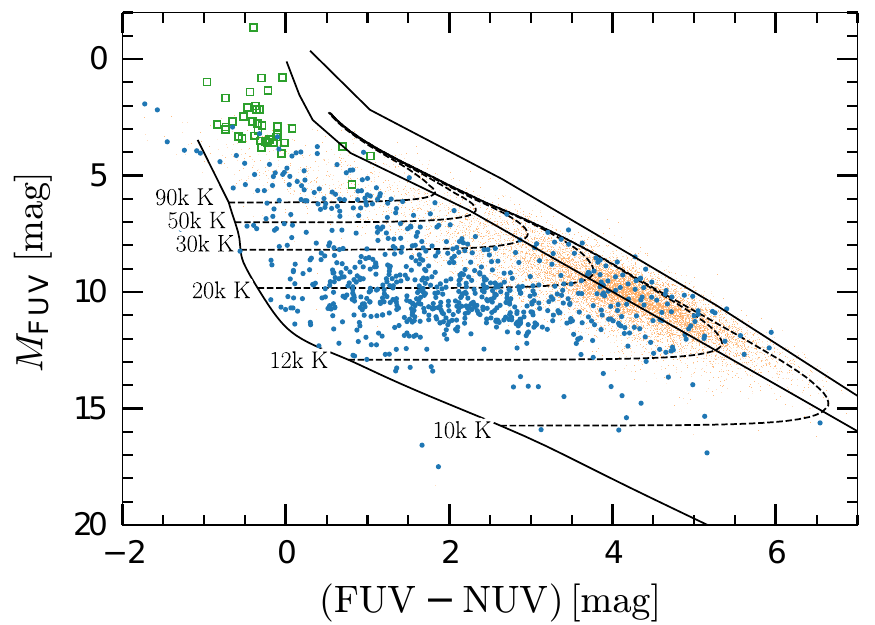} 
   \caption{\emph{GALEX} HR diagram. Symbols and colours as in Figure\,\ref{fig:hr_tgas_dr2}  for the TGAS-\emph{GALEX} cross-matched stars and our 775 WD+AFGK candidates. Hot-subdwarf with AFGKM companions \citep{Geier2020A&A...635A.193G} are plotted as green squares. The intrinsic magnitudes and colours of $\log{g} = 8$ WDs are shown on the left as solid black curve, while the composite colours and absolute magnitudes of WD+AFGK binaries are plotted as dashed curves. The Solar-metallicity ZAMS and TAMS are plotted as in Figure\,\ref{fig:hr_tgas_dr2}.}
   \label{fig:hr_galex}
\end{figure}

\subsection{Cross-match with the second data release of Gaia}

Two years after the release of TGAS, the second data release of \emph{Gaia} made available $G_\mathrm{BP}$ and $G_\mathrm{RP}$ photometry and improved astrometry for all our previously selected binary candidates. The new data caused a significant improvement, of which we have taken advantage for the refinement of the TGAS WD+AFGK candidate selection via the SED-fitting procedure described in the previous section. We adopted the {\em Gaia} DR2 parallax zero-point of $-0.029$\,mas \citep{Lindegren2018A&A...616A...2L} and the revised passbands, $G$-band correction, and zero-points \citep{Maiz-apellaniz2018A&A...619A.180M}. We also used new 2MASS zero-points \citep{Maiz-apellaniz2018A&A...616L...7M}. Because of the high-precision of the {\em Gaia}-DR2 parallaxes, adopting $d \propto \varpi^{-1}$ is an accurate approximation of the distance-estimates provided by \citet{Bailer2018AJ....156...58B}. This refined sample contains 814 WD+AFGK candidates (Figure\,\ref{fig:colorcut}). In addition, we estimated the uncertainties of photometrically determined stellar parameters by using the \textsc{python} Markov Chain Monte Carlo sampler \citep[\textsc{emcee};][]{Foreman2019JOSS....4.1864F}. 
The Appendix Table\,\ref{tab:tgas-wdfgk-sample} contains the relevant photometric and astrometric data, as well as the results of the SED-fitting analysis.

The striking improvement delivered by {\em Gaia} DR2 is visualized through the comparison between the TGAS and \emph{Gaia} DR2 Hertzsprung-Russell (HR) diagrams in Figure\,\ref{fig:hr_tgas_dr2}, where main-sequence and giant stars can be unequivocally identified (as noted in Table\,\ref{tab:tgas-wdfgk-sample}). In addition, the \emph{Gaia} DR2 parallaxes can also be used to construct a \emph{GALEX} HR diagram, showing that the contamination of our sample from other compact UV-bright objects, like the hot-subdwarf stars \citep{Geier2020A&A...635A.193G}, is minimal (see Figure\,\ref{fig:hr_galex}).

\subsection{Cross-match with SIMBAD}

We cross-match the 814 selected WD+AFGK candidates with SIMBAD \citep{Wenger2000A&AS..143....9W}, to obtain their SIMBAD classifications. Most of our systems are classified as ``Star" or ``High proper-motion Star". There are also 36 WD+AFGK candidates that are classified as ``Carbon star" (1), ``Classical Cepheid" (1), ``CV DQ Her type" (1), ``Eclipsing binaries" (19), ``Eruptive variable star" (2), ``Galaxy" (1), ``Herbig Ae/Be star" (1), ``Horizontal branch star" (1), ``Hot subdwarf" (2), ``Nova-like star" (1), ``Planetary nebula" (1), ``Pulsating variable star" (2), ``T Tau-type star" (1), ``Variable star of RR Lyr type" (2). We exclude these 36 ``possible contaminants" from our WD+AFGK sample, which is thus reduced to 778 systems. 

Furthermore, of the 778 system, there are three candidates (TYC\,3793-959-1, TYC\,265-1112-1, and TYC\,2523-2620-1) which are re-classified as contaminants (MS binaries or triple systems harboring two MS stars and a WD) by us later after investigating our follow-up high resolution spectra (displaying double-lines), which will be published in a forthcoming paper. Thus, there are in total 39 ``possible contaminants" in our sample. After excluding them, we finally obtain 775 WD+AFGK candidates. The SIMBAD classification information of the 775 WD+AFGK candidates and 39 excluded ``possible contaminants" are listed in Table\,\ref{tab:tgas-wdfgk-sample}. 

\subsection{Cross-match with RAVE DR5 and LAMOST DR6}

RAVE is a multi-fibre spectroscopic astronomical survey of stars in the Milky Way. It is operated through the 1.2-m UK Schmidt Telescope of the Australian Astronomical Observatory (AAO). As a Southern hemisphere survey, RAVE covers 20\,000 square degrees of the sky. The primary aim of RAVE is to derive the RVs, stellar parameters and elemental abundances of stars to study the structure, formation and evolution of our Milky Way \citep{Kunder2017AJ....153...75K}. RAVE targets bright stars in the magnitude range 8\,$<$\,I\,$<$\,12\,mag. The RAVE spectra cover the spectral region 8410\,--\,8794\,\AA\ which contains the infrared Calcium triplet, with a resolving power of R\,$\sim$\,7500. This allows obtaining RVs with a median precision better than 1.5\,\kms\ and good precision stellar atmospheric parameters and chemical abundances \citep{Kunder2017AJ....153...75K}. 
RAVE largely overlaps with TGAS, i.e. 249\,603 spectra of 215\,590 unique TGAS stars have been observed by RAVE. By comparing our WD+AFGK list with the newest release of RAVE \cite[i.e. DR5;][]{Kunder2017AJ....153...75K}, we found 104 systems in common (corresponding to 125 RAVE spectra, marked in the ``Spec" column in Table\,\ref{tab:tgas-wdfgk-sample}).

 LAMOST is a quasi-meridian reflecting Schmidt telescope of $\sim$\,4\,m effective aperture and a field of view of 5\,deg in diameter \citep{Cui2012RAA....12.1197C, Liu2020arXiv200507210L}, located in Xinglong station of National Astronomical Observatories, Chinese Academy of Sciences. Being a dedicated large-scale survey telescope, LAMOST uses 4000 fibres to obtain spectra of celestial objects as well as sky background and calibration sources in one single exposure. LAMOST spectra cover the entire optical wavelength range ($\simeq$\,3700\,--\,9000\AA) at a resolving power R $\sim$ 1800 \citep{Luo2015RAA....15.1095L}. LAMOST DR6 low resolution spectroscopic survey has made use of 4577 plates, of which 75\,\% are VB/B plates (very bright plates: r\,$\leq$\,14\,mag, bright plates: 14\,mag\,$\leq$\,$r$\,$\leq$\,16.8\,mag), 41\,\% are VB plates, similar to those summarized in \citet{Ren2018MNRAS.477.4641R}. Thus, a large fraction of stars observed by LAMOST are bright. By cross-matching our WD+AFGK binary list with LAMOST DR6, we found 82 targets in common (corresponding to 138 LAMOST spectra, see the ``Spec" column in Table\,\ref{tab:tgas-wdfgk-sample}).

Table\,\ref{tab:summary_sample} presents a summary of our WD+AFGK sample. As mentioned in Section 2.3, after excluding the 37 ``possible contaminants", we obtain a sample containing 775 WD+AFGK candidates. Of them, 37 and 20 candidates have been published in our previous RAVE DR4 \citepalias{Parsons2016MNRAS.463.2125P} and LAMOST DR4 WD+AFGK sample \citepalias{Rebassa2017MNRAS.472.4193R}, respectively. The last column of Table\,\ref{tab:tgas-wdfgk-sample} marks those already published before. Therefore, 718 are new WD+AFGK candidates, of which 67 have the new available RAVE DR5 spectra and 62 have the LAMOST DR6 spectra (see Appendix Table\,\ref{tab:param_tgas-rave-lamost} for the detailed information), all of which will be used in our following RV measurements. 

\begin{deluxetable}{lr}
\setlength{\tabcolsep}{3pt}
\tablenum{1}
\tablecaption{Summary of the WD+AFGK candidates identified in this work. \label{tab:summary_sample}}
\tabletypesize{\scriptsize}
\tablewidth{0pt}
\tablehead{
\colhead{Name} & \colhead{Number}
}
\startdata
Total number of sources with UV colors and photometric \Teff & $\approx$\,13\,000 \\
The initial selected WD+AFGK candidates in TGAS & 985 \\
The \emph{Gaia} DR2 refined sample of WD+AFGK candidates & 814 \\
The final sample after removing contaminants  & 775 \\
\ \ \ \ with available RAVE DR5 spectra           & 104 \\
\ \ \ \ with available LAMOST DR6 spectra         & 82 \\
    \enddata
\end{deluxetable}

\begin{deluxetable*}{lcccccc}
\tablenum{2}
\tablecaption{The observation summary of high resolution spectroscopy. \label{tab:obs_summary}}
\tablewidth{0pt}
\tablehead{
\colhead{Date} & \colhead{N$_\mathrm{Spec}$} & \colhead{Telescope} & \colhead{Weather} & \colhead{Seeing} & \colhead{Exposure time} & \colhead{S/N} \\
\colhead{} & \colhead{} &  \colhead{} & \colhead{} & \colhead{(arcsec)} & \colhead{(second)} & \colhead{}
}
\startdata
20170706 &  3 & SPM$+$Echelle & cirrus       & 1.6 & $\simeq$\,1200   & 5\,--\,10  \\
20170709 &  3 & SPM$+$Echelle & cirrus       & 1.8 & $\simeq$\,1200   & 5\,--\,10  \\
20170710 &  1 & SPM$+$Echelle & cirrus       & 1.6 & $\simeq$\,1200   & 13         \\
20170711 &  1 & SPM$+$Echelle & cirrus       & 1.6 & $\simeq$\,1200   & 13         \\
20171202 & 18 & XL216$+$HRS   & clear        & 2.3 & 100\,--\,2400  & 25\,--\,100\\
20171203 & 15 & XL216$+$HRS   & clear        & 2.7 & 400\,--\,3300  & 25\,--\,80 \\
20171204 & 18 & XL216$+$HRS   & clear        & 2.5 & 200\,--\,2400  & 25\,--\,80 \\
20180104 &  9 & XL216$+$HRS   & cloudy       & 2.5 & 400\,--\,3600  & 25\,--\,110\\
20180110 & 10 & SPM$+$Echelle & cloudy/windy & 2.1 & $\simeq$\,1200   & 10\,--\,20 \\
20180110 &  4 & XL216$+$HRS   & clear        & 3.5 & 360\,--\,3600  & 25\,--\,40 \\
20180111 &  3 & SPM$+$Echelle & cloudy/windy & 1.8 & $\simeq$\,1200   & 10\,--\,22 \\
20180111 & 19 & XL216$+$HRS   & clear        & 2.5 & 100\,--\,1200  & 10\,--\,50 \\
20180112 & 10 & SPM$+$Echelle & cloudy/windy & 1.8 & $\simeq$\,1200   & 12\,--\,19 \\
20180113 &  9 & SPM$+$Echelle & cirrus/windy & 1.8 & $\simeq$\,1200   & 9\,--\,16  \\
20180114 &  5 & SPM$+$Echelle & clear        & 1.3 & $\simeq$\,1200   & 11\,--\,21 \\
20180116 &  5 & SPM$+$Echelle & cirrus       & 1.8 & $\simeq$\,1200   & 9\,--\,18  \\
20180220 & 19 & XL216$+$HRS   & cloudy       & 2.0 & 40\,--\,2400   & 25\,--\,90 \\
20180221 & 15 & XL216$+$HRS   & clear        & 2.0 & 500\,--\,2400  & 30\,--\,70 \\
20180222 & 13 & XL216$+$HRS   & cloudy       & 2.0 & 100\,--\,2400  & 25\,--\,65 \\
20180225 &  3 & XL216$+$HRS   & cloudy       & 2.0 & 900\,--\,2400  & 25\,--\,80 \\
20180226 & 15 & XL216$+$HRS   & clear        & 2.0 & 400\,--\,1400  & 25\,--\,45 \\
20180305 &  3 & XL216$+$HRS   & clear        & 4.0 & 3600           & 27\,--\,37 \\
20180502 &  9 & XL216$+$HRS   & clear        & 2.3 & 200\,--\,1800  & 22\,--\,50 \\
20180503 & 13 & XL216$+$HRS   & clear        & 2.0 & 180\,--\,1400  & 14\,--\,50 \\
20180504 &  4 & XL216$+$HRS   & cloudy       & 2.0 & 1200\,--\,3000 & 20\,--\,30 \\
20180505 &  5 & XL216$+$HRS   & cloudy       & 2.2 & 300\,--\,3300  & 20\,--\,40 \\
20180506 &  8 & XL216$+$HRS   & cloudy       & 2.0 & 900\,--\,3000  & 20\,--\,37 \\
20180527 &  4 & XL216$+$HRS   & cloudy       & 2.1 & 1800\,--\,2400 & 15\,--\,39 \\
20180601 &  6 & XL216$+$HRS   & clear        & 2.1 & 400\,--\,2400  & 29\,--\,40 \\
20180602 &  3 & XL216$+$HRS   & cloudy       & 2.2 & 1000\,--\,3600 & 22\,--\,49 \\
20180607 &  6 & XL216$+$HRS   & clear        & 2.5 & 30\,--\,1500   & 28\,--\,65 \\
20180626 &  2 & XL216$+$HRS   & clear        & 2.5 & 1800\,--\,3600 & 35\,--\,40 \\
20180627 &  3 & XL216$+$HRS   & clear        & 2.1 & 1200\,--\,1800 & 40\,--\,46 \\
\enddata
\tablecomments{In this table we list the observational date, the number of spectra obtained per night, the instruments (telescopes and spectrographs) we used, the corresponding weather, seeing, exposure time range and S/N range (S/N is estimated at $\sim$\,8500\,\AA\ for XL216+HRS spectrum, and $\sim$\,6000\,\AA\ for SPM+Echelle spectrum) during the observations.}
\end{deluxetable*}

\section{Observations}
\label{sec:observations}

As demonstrated in \citetalias{Rebassa2017MNRAS.472.4193R}, high-resolution spectra are needed to efficiently identify low inclination ($\lesssim$\,5\,deg) and/or long orbital period systems ($\gtrsim$\,100\,d), which is especially important for identifying systems that will evolve through a second CE phase and thus become SN Ia double-degenerate progenitor candidates. Our observations were hence carried out by using the Xinglong 2.16\,m telescope \citep[hereafter, XL216;][]{Fan2016PASP..128k5005F} and the San Pedro M\'artir 2.12\,m telescope, both equipped with Echelle spectrographs.

We observed 93 WD+AFGK candidates during 23 nights from the XL216. The observing log is summarized in Table\,\ref{tab:obs_summary}. The instrument terminal we used was the High Resolution fiber-fed Spectrograph (HRS), thus providing Echelle spectra of a 49\,800 resolving power for a fixed slit width of 0.19\,mm, and covering the $\sim$\,3650\,--\,10\,000\,\AA\ wavelength range \citep{Fan2016PASP..128k5005F}. The Thorium-Argon (hereafter, Th-Ar) arc spectra were taken at the beginning and end of each night. The HRS works in a very stable environment, where the temperature is quite stable. The stability of the HRS instrument for the RV measurement is rms\,=\,$\pm$\,6\,m\,s$^{-1}$ \citep{Fan2016PASP..128k5005F}. 214 spectra were obtained for 93 WD+AFGK candidates, of which 92 were observed at least twice and on different nights. Only one system (TYC\,4698-895-1) was observed just once. The spectra were reduced by using the IRAF package \citep{Tody1986SPIE..627..733T, Tody1993ASPC...52..173T} following the standard procedures: bias subtraction, flat-field correction, scattered-light subtraction, spectra extraction, wavelength calibration (based on Th-Ar arc spectra), and continuum normalization (by using the IRAF task ``continuum" order by order).

Eight nights of observations were carried out using the Echelle spectrograph attached to the 2.12\,m telescope at the San Pedro M\'artir Observatory in Baja California, M\'exico. The corresponding resolving power is $R$\,$\sim$\,20\,000 for a slit width of 2.8\,arcsec, covering the 3650\,--\,7300\,\AA\ wavelength range. Arc spectra were taken before and after each object. 50 high resolution spectra were obtained for 22 WD+AFGK candidates, of which 18 have at least two spectra obtained on different nights. The data reduction procedures of SPM spectra are the same as the XL216 spectra.

To summarize, we obtained 264 high resolution spectra for 104 WD+AFGK candidates, each of them having at least two spectra obtained on different nights.

\begin{figure}
   \centering
   \includegraphics[width=0.485\textwidth]{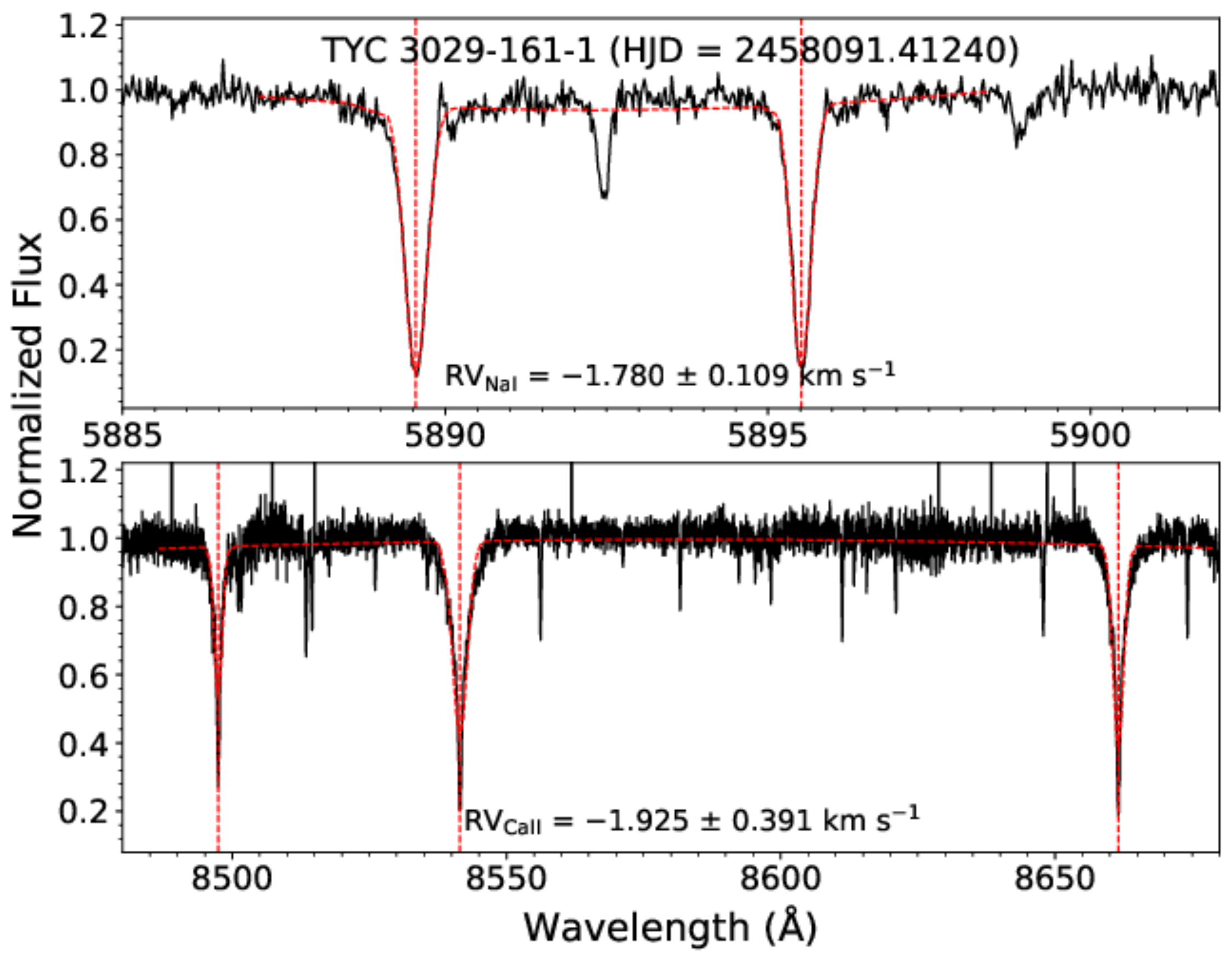} 
   \caption{Line fitting examples for the \Ion{Na}{I}\,D doublet (top panel) and \Ion{Ca}{II} triplet\ (bottom panel) profile of a high resolution spectrum from XL216 telescope. The vertical red dashed line shows the fitted line center. The fitted RVs and fitting errors are shown in the figure too.}
   \label{fig:rv_example}
\end{figure}

\section{Radial Velocity measurements}

\subsection{High resolution spectra}

We measured the RVs from the 214 spectra obtained from the XL216 telescope by fitting the normalised \Ion{Ca}{II} absorption triplet (at 8498.03, 8542.09, and 8662.14\,\AA) with a combination of a second order polynomial and a triple-Gaussian profile of fixed separations, as described in \citetalias{Rebassa2017MNRAS.472.4193R}. Only when the \Ion{Ca}{II} absorption triplet was too noisy to get a reliable RV, the normalized \Ion{Na}{I} doublet at $\sim$\,5890\,\AA\ (i.e. 5889.951 and 5895.924\,\AA) was used. In this case we used a second order polynomial and a double-Gaussian profile of fixed separation. The RV uncertainty is obtained by summing the fitted error and a systematic error of 0.5\,\kms in quadrature, which is an appropriate value for spectra of Signal-to-Noise ratio (S/N)\,$\sim$\,25--30, based on our experience with this instrument.

\begin{figure}
   \centering
   \includegraphics[width=0.480\textwidth]{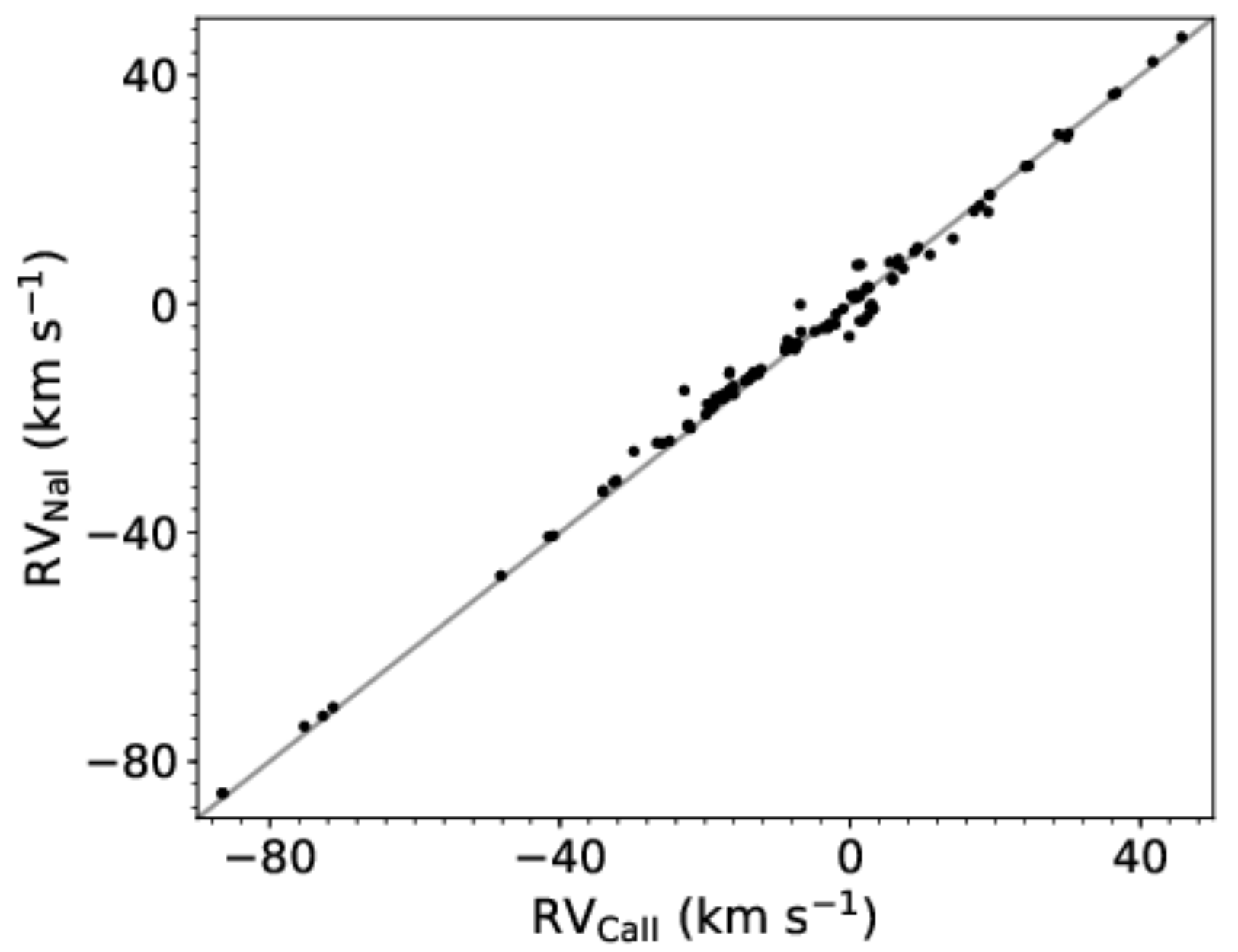} 
   \caption{Comparison of RVs determined by fitting \Ion{Ca}{II} triplet and \Ion{Na}{I} doublet for the 112 spectra observed with the XL216.}
   \label{fig:rv_comparison}
\end{figure}

An example of double Gaussian fit to the \Ion{Na}{I} doublet profile and triple Gaussian fit to the \Ion{Ca}{II} triplet profile can be seen in Figure\,\ref{fig:rv_example}. Figure\,\ref{fig:rv_comparison} shows a comparison between the RVs determined from \Ion{Ca}{II} triplet and the \Ion{Na}{I} doublet from the same spectra, which are in good agreement despite the fact that the \Ion{Na}{I} feature is usually affected by interstellar absorption.

Given that the SPM spectra only cover the 3650\,--\,7300\,\AA\, wavelength range, the RVs are determined by fitting the \Ion{Na}{I} doublet. The RV uncertainty is obtained by summing up the error obtained from the fit and a systematic error of 1\,\kms\ ($R$\,$\sim$\,20\,000) in quadrature.

\subsection{RVs from RAVE/LAMOST spectra}

The RVs of the 104 WD+AFGK candidates with 125 RAVE spectra are taken from the DR5 catalogue \citep{Kunder2017AJ....153...75K}, which are determined by an automatic pipeline using a standard cross-correlation procedure. For the RAVE RV uncertainties, we incorporate a systematic error of 3\,\kms\ due to the medium resolution ($R$\,$\sim$\,7500) of RAVE spectra. For the 138 LAMOST spectra of 82 candidates as mentioned in Section\,2, We measured the RVs for 128 good quality spectra of 80 WD+AFGK candidates by fitting the \Ion{Ca}{II} absorption triplet, the other 10 spectra have too bad quality to measure RVs. For the LAMOST RV uncertainties, the systematic error we incorporate is 10\,\kms\ \citep{Luo2015RAA....15.1095L}.

\begin{deluxetable}{lrrrrc}
\setlength{\tabcolsep}{1pt}
\tablenum{3}
\tablecaption{The RVs table of the 275 WD+AFGK candidates. \label{tab:rv}}
\tabletypesize{\scriptsize}
\tablewidth{0pt}
\tablehead{
\colhead{Name} & \colhead{HJD} & \colhead{RV}      & \colhead{Err$_\mathrm{fit}$}  & \colhead{Err$_\mathrm{tot}$}  & \colhead{Telescope}  \\
\colhead{}   & \colhead{(d)} &  \colhead{(\kms)} &  \colhead{(\kms)} &  \colhead{(\kms)} & \colhead{}         
}
\startdata
TYC\,1006-4-1    & 2458170.36445 & $-$18.854 & 0.419 &  0.652 & XL216  \\
TYC\,1006-4-1    & 2458172.30624 & $-$19.139 & 0.469 &  0.685 & XL216  \\
TYC\,1010-403-1  & 2457860.34272 & $-$16.685 & 5.696 & 11.509 & LAMOST \\
TYC\,1010-403-1  & 2457917.19322 & $-$26.570 & 4.586 & 11.001 & LAMOST \\
TYC\,1020-875-1  & 2457528.30082 & $-$18.835 & 2.986 & 10.436 & LAMOST \\
TYC\,110-755-1   & 2458170.07184 & $-$37.246 & 0.255 &  0.561 & XL216  \\
TYC\,110-755-1   & 2458176.04568 & $-$42.918 & 0.589 &  0.772 & XL216  \\
TYC\,1131-1838-1 & 2458272.28582 &  $-$7.703 & 0.334 &  0.601 & XL216  \\
TYC\,1131-1838-1 & 2458277.31122 &  $-$7.257 & 0.241 &  0.555 & XL216  \\
TYC\,1191-179-1  & 2457941.89752 &    10.143 & 0.477 &  1.108 & SPM    \\
\nodata          & \nodata       & \nodata   & \nodata & \nodata & \nodata \\
\enddata
\tablecomments{This table also lists the HJD, and the telescopes (i.e. XL216, SPM, LAMOST, RAVE) used for obtaining the spectra.}
\end{deluxetable}

\subsection{The final RV table}

By including all the RVs determined from the previous subsections, we have obtained 517 RV values for 275 WD+AFGK candidates. All the RVs are listed in Table\,\ref{tab:rv}, which presents the corresponding Heliocentric Julian Dates (HJD), the RVs and corresponding errors (fitting and total errors), and the telescopes used for obtaining the spectra. In summary, for the 275 WD+AFGK candidates with available RVs, 154 targets have at least two RVs, 151 have at least two RVs separated by one night (i.e. only three targets have two RVs at same night), and 121 have only one RV value (which are either from LAMOST or RAVE). Most importantly, for the 264 RVs (104 targets) obtained from XL216/SPM high resolution spectral follow-up, all the 104 targets have at least two RVs separated by one night. Table\,\ref{tab:stat_RVs} presents the statistics of all the 517 RVs from different telescopes.

\begin{deluxetable}{cccc}
\setlength{\tabcolsep}{3pt}
\tablenum{4}
\tablecaption{The number statistics of available RVs of WD+AFGK candidates from different telescopes. \label{tab:stat_RVs}}
\tabletypesize{\scriptsize}
\tablewidth{0pt}
\tablehead{
\colhead{Telescope} & \colhead{$\mathrm{N_{RVs}}$}    & \colhead{$\mathrm{N_{target}}$} & \colhead{$\mathrm{N^{targets} _{\geq\,2\,RVs}}$}  
}
\startdata
XL216     & 214  & 93                 & 92  \\
SPM       & 50   & 22                 & 18  \\
XL216+SPM & 264  & 104$^\mathrm{[a]}$ & 104 \\
RAVE      & 125  & 104$^\mathrm{[b]}$ & 20  \\
LAMOST    & 128  & 80$^\mathrm{[c]}$  & 29  \\
Total     & 517  & 275                & 151 \\
\enddata
\tablecomments{This table shows the number statistics of available RVs for different RV origins for 275 WD+AFGK candidates, which includes the Telescopes, number of RVs, the corresponding target number, and the number of targets with at least two RVs separated by one night. [a]: Note that 11 targets were both observed by XL216 and SPM. [b]: One of them was also observed by XL216, and one by SPM. [c]: 9 of them were also observed by XL216, 2 by SPM.}
\end{deluxetable}

\subsection{Confirmed Radial Velocity Variables}

We use the RVs from Table\,\ref{tab:rv} to identify close binaries. That is, a given system will be considered as a close binary if we detect significant (i.e. $>$\,3\,$\sigma$) RV variation \citep{Rebassa2007MNRAS.382.1377R, Ren2013AJ....146...82R}. Conversely, if no RV variation is detected from spectra observed in at least two different nights, the system is considered as a likely wide binary candidate. 

We identify 23 RV variable AFGK stars (displaying more than 3\,$\sigma$ RV variation), which are suggested to be members of close binary systems; the remaining 128 targets are suggested to be likely wide-binary members. Table\,\ref{tab:stat_close_wide} gives the detailed number of close/wide WD+AFGK candidates identified from different telescopes. 

\begin{figure}
   \centering
   \includegraphics[width=0.480\textwidth]{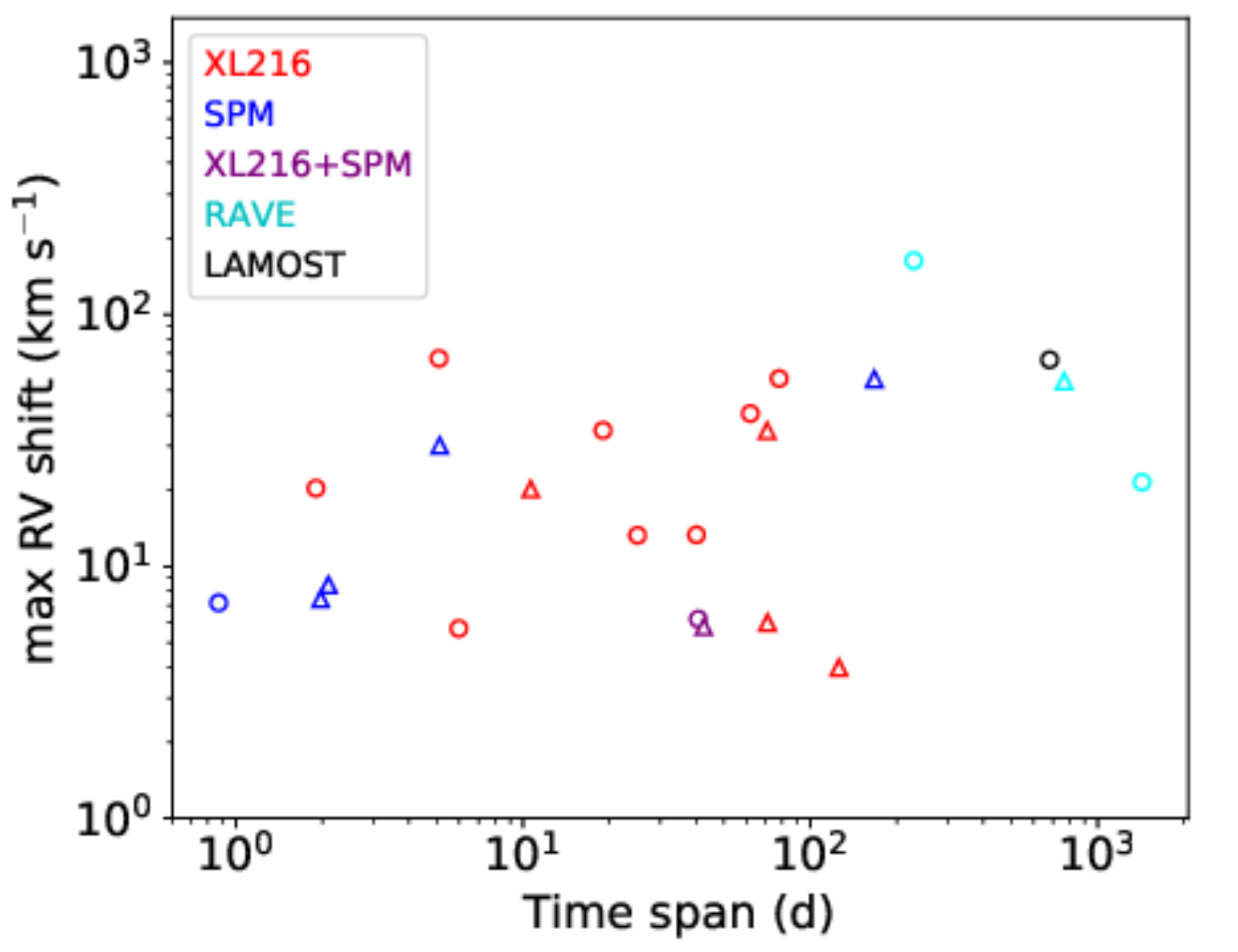} 
   \caption{The maximum RV shift versus time span for the 23 close binaries identified in this work. The circles show the dwarfs (13), while the triangles show the giants (10). Different colors show the close systems detected from different telescopes.}
   \label{fig:rv_shift}
\end{figure}

\begin{deluxetable}{ccccc}
\setlength{\tabcolsep}{3pt}
\tablenum{5}
\tablecaption{The number of close/wide system candidates detected from different telescopes in this work. \label{tab:stat_close_wide}}
\tabletypesize{\scriptsize}
\tablewidth{0pt}
\tablehead{
\colhead{Type} & \colhead{XL216/SPM} & \colhead{RAVE} & \colhead{LAMOST} & \colhead{Total} 
}
\startdata
Close       & 19  & 3  & 1  & 23 \\
Wide        & 85  & 17 & 26 & 128  \\
Close+Wide  & 104 & 20 & 27 & 151  \\
\enddata
\tablecomments{This table shows the statistics of close/wide system candidates detected from different telescopes. }
\end{deluxetable}

\begin{deluxetable}{lrrc}
\setlength{\tabcolsep}{2pt}
\tablenum{6}
\tablecaption{The information of the 23 close binaries identified in this work. \label{tab:table_close}}
\tabletypesize{\scriptsize}
\tablewidth{0pt}
\tablehead{
\colhead{Name} & \colhead{max RV shift} & \colhead{Time span} & \colhead{Detect from} \\
\colhead{} & \colhead{(\kms)} & \colhead{(d)} & \colhead{} 
}
\startdata
TYC\,110-755-1   &   5.672 &    5.97384 & XL216     \\
TYC\,1223-498-1  &  20.254 &   10.63028 & XL216     \\
TYC\,1380-957-1  &  66.925 &    5.09723 & XL216     \\
TYC\,1394-1008-1 &  13.350 &   40.14194 & XL216     \\
TYC\,1428-81-1   &  30.225 &    5.12495 & SPM       \\
TYC\,1655-707-1  &  34.656 &   18.97566 & XL216     \\
TYC\,2292-1379-1 &   6.178 &   40.73237 & XL216+SPM \\
TYC\,278-239-1   &   8.445 &    2.09796 & SPM       \\
TYC\,2850-1366-1 &  20.429 &    1.90052 & XL216     \\
TYC\,3104-932-1  &  13.294 &   25.02232 & XL216     \\
TYC\,3814-455-1  &  66.084 &  681.05243 & LAMOST    \\
TYC\,3883-1104-1 &   3.964 &  125.86194 & XL216     \\
TYC\,418-2364-1  &   5.987 &   70.98452 & XL216     \\
TYC\,4564-627-1  &  40.447 &   61.80570 & XL216     \\
TYC\,4700-815-1  &  55.650 &   77.90281 & XL216     \\
TYC\,4717-255-1  &   5.763 &   42.55069 & XL216+SPM \\
TYC\,5523-324-1  &   7.167 &    0.87083 & SPM       \\
TYC\,5856-1958-1 &  21.590 & 1429.05347 & RAVE      \\
TYC\,7443-1018-1 &  54.294 &  765.94586 & RAVE      \\
TYC\,841-433-1   &  34.559 &   70.82929 & XL216     \\
TYC\,856-918-1   &   7.422 &    1.97168 & SPM       \\
TYC\,8873-148-1  & 163.580 &  229.34639 & RAVE      \\
TYC\,969-1420-1  &  55.438 &  167.17345 & SPM       \\
\enddata
\tablecomments{This table lists the maximum RV shift and the corresponding time span for the 23 close binaries. The telescopes used to detect the close binaries are also listed.}
\end{deluxetable}

Table\,\ref{tab:table_close} lists the 23 close binaries and the corresponding telescopes used to detect them, as well as the maximum RV shift measured and the corresponding time span between the observations. Figure\,\ref{fig:rv_shift} plots the maximum RV shift versus time span for the 23 close binaries. We can see that the maximum RV shifts vary from $\sim$\,4 to 160\,\kms. The close binaries with RV shifts as small as 4\,\kms were detected from our highest resolution ($R$\,$\sim$\,49\,800) spectra (i.e. XL216 data, red open circles in Figure\,\ref{fig:rv_shift}), which agree with the statement of \citetalias{Rebassa2017MNRAS.472.4193R} claiming that high resolution spectra are needed to identify close binaries with small RV variations (i.e. these systems have long orbital periods and/or low orbital inclinations; see Figure\,5 in \citetalias{Rebassa2017MNRAS.472.4193R}). Furthermore, most of our close binaries have the time baseline shorter than $\sim$\,100\,d, as our follow-up observations were performed within half a year. Only four systems detected from survey data (i.e. LAMOST/RAVE) have considerably longer time baselines.

\section{Stellar Parameters}
\label{sec:classification}

\subsection{Dwarf/Giant classification}

When selecting TGAS WD+AFGK binaries, we only used the \Teff\ vs. FUV$-$NUV diagram, without applying a \logg\ cut. Thus, our WD+AFGK binary sample contains both AFGK dwarfs and giants. 

The dwarf/giant classification is based on the \emph{Gaia} DR2 HR diagram. As inferred from the bottom panel of Figure\,\ref{fig:hr_tgas_dr2}, the stars below/above the TAMS are classified as dwarfs/giants respectively. The dwarf/giant classification is flagged in the Appendix Table\,\ref{tab:tgas-wdfgk-sample}. Of the 775 WD+AFGK binaries that form our sample, 443/332 are classified as dwarfs/giants, i.e. a giant fraction $\sim$\,43\%. Among the 23 close WD+AFGK candidates we identified (as shown in Table\,\ref{tab:table_close}), 10 are giants, which corresponds to a giant fraction $\sim$\,43\% for close systems. We will discuss the close binary fractions of WD+AFGK binaries containing dwarf and giant stars in Section 6.

\subsection{Stellar parameters from RAVE/LAMOST spectra}

For those WD+AFGK binaries with available RAVE DR5 medium-resolution spectra, the stellar atmospheric parameters and chemical abundances of their companions were adopted from \citet{Kunder2017AJ....153...75K}, who used the same stellar parameter pipeline as in DR4, but calibrated using recent K2 Campaign 1 seismic gravities and $Gaia$ benchmark stars, as well as results obtained from high-resolution studies. The typical uncertainties in $\Teff$, $\logg$ and [M/H] are approximately 250\,K, 0.4\,dex and 0.2\,dex respectively, but vary with stellar population and S/N. The stellar parameters of the 104 WD+AFGK binaries (125 RAVE spectra) are listed in Table\,\ref{tab:param_tgas-rave-lamost} in the Appendix. 

The stellar parameters of the TGAS-LAMOST WD+AFGK candidates were determined from the LAMOST Stellar Parameter pipeline \citep[LASP;][]{Luo2015RAA....15.1095L}. LASP determines the stellar parameters by template matching with the ELODIE empirical library \citep{Prugniel2001A&A...369.1048P}. The stellar parameters of the AFGK companions of 82 WD+AFGK binaries are listed in Table\,\ref{tab:param_tgas-rave-lamost} of the Appendix too.

\begin{figure}
   \centering
   \includegraphics[width=0.45\textwidth]{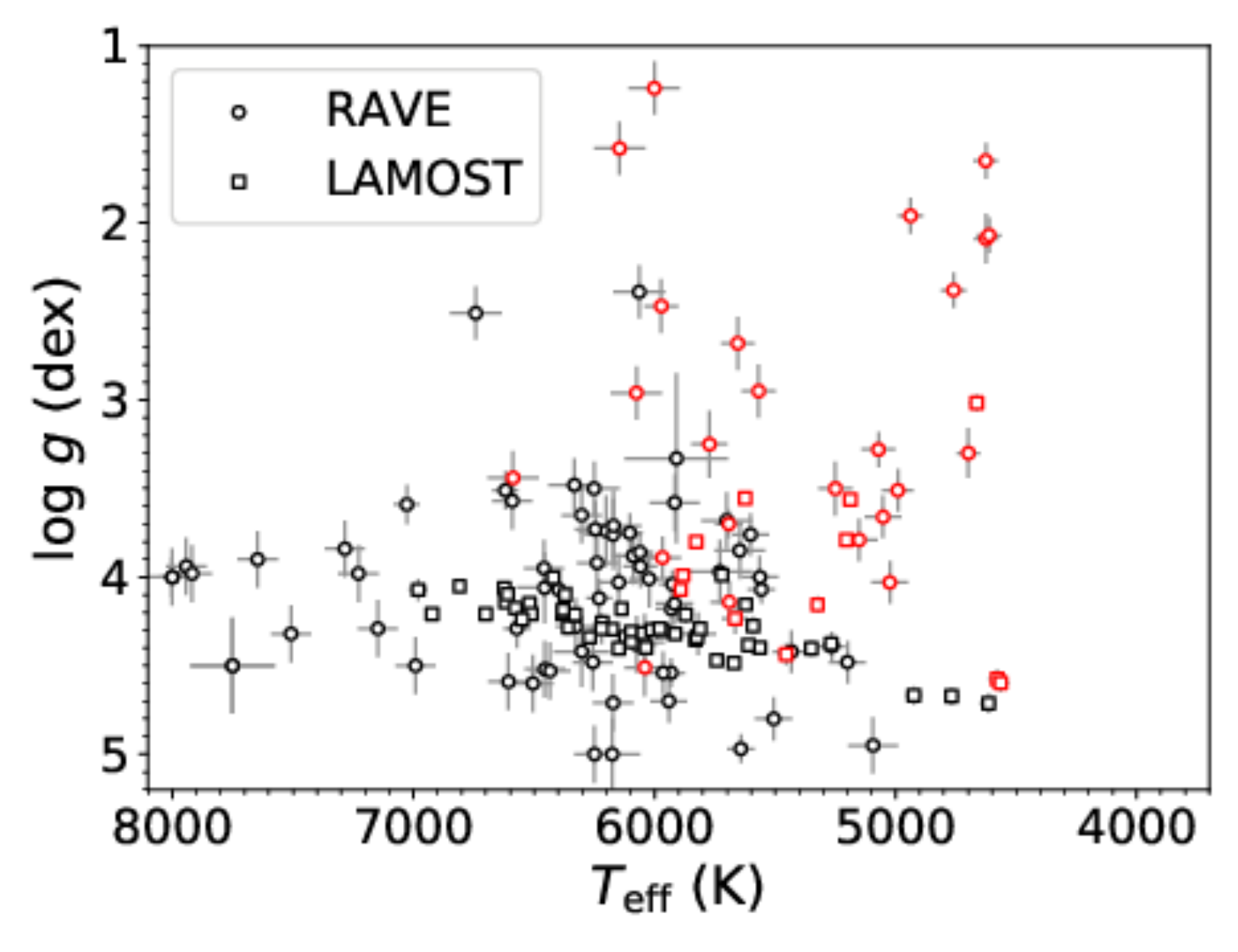} 
   \caption{The \logg\ vs. \Teff\ diagram for TGAS-RAVE/LAMOST WD+AFGK candidates with available RAVE/LAMOST stellar parameters (for S/N\,$>$\,30). The circle/square marks the  stellar parameters from RAVE/LAMOST respectively. The black/red colors correspond to the dwarf/giant classification of Section 5.1.}
   \label{fig:hr_rave-lamost}
\end{figure}

\begin{figure}
   \centering
   \includegraphics[width=0.45\textwidth]{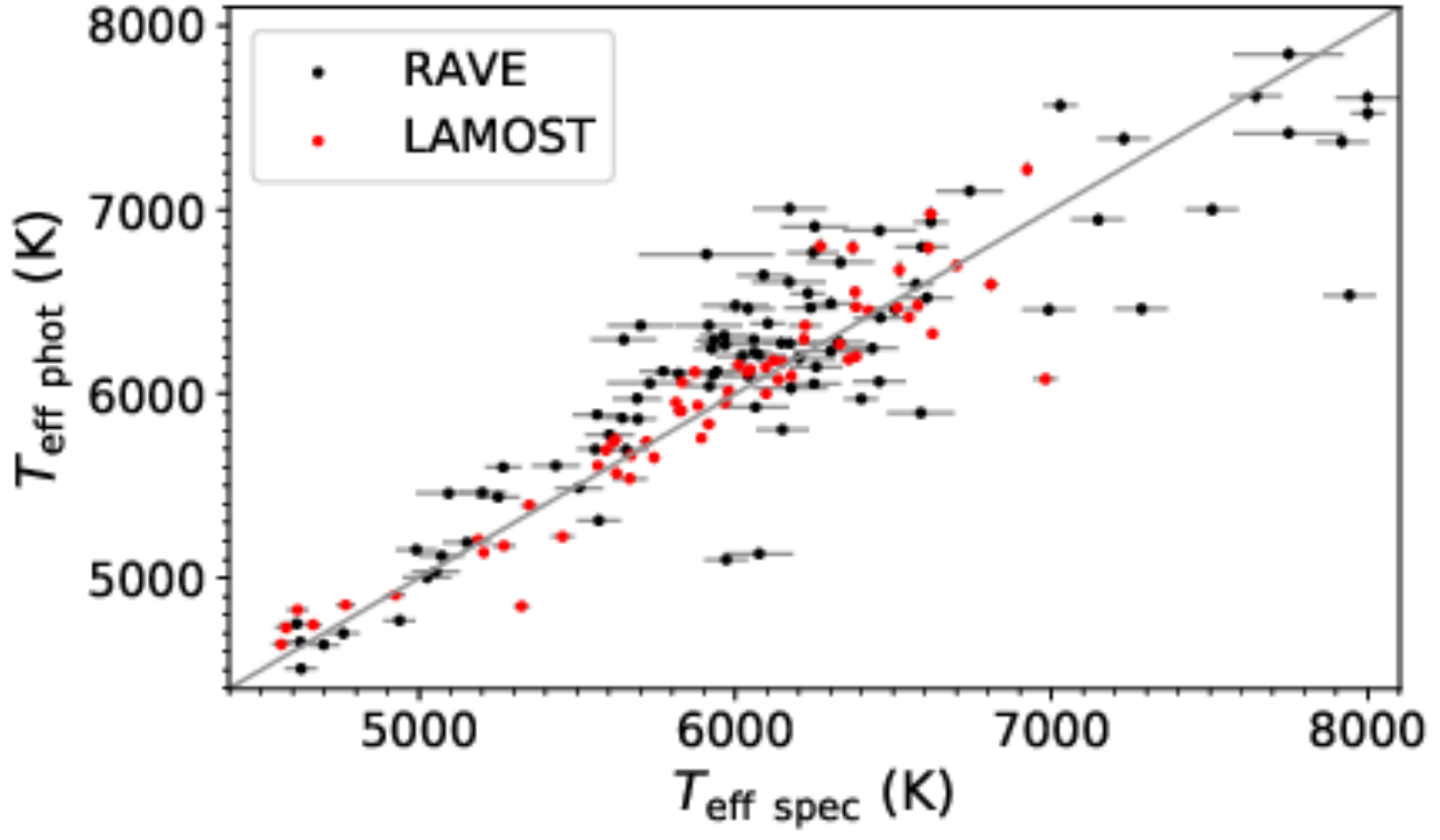} 
   \caption{The comparison of spectroscopic \Teff\ determined from RAVE/LAMOST spectra and the photometric \Teff\ from Table\,\ref{tab:tgas-wdfgk-sample}. The black/red colors identify the RAVE/LAMOST samples, respectively.}
   \label{fig:compare_teff_lamost-rave}
\end{figure}

In brief, of the 186 WD+AFGK binaries with available RAVE/LAMOST spectra shown in Table\,\ref{tab:param_tgas-rave-lamost}, 154 have available RAVE/LAMOST stellar parameters and S/N\,$>$\,30. Their \logg\ vs. \Teff\ diagram is shown in Figure\,\ref{fig:hr_rave-lamost}. The black/red dots in Figure\,\ref{fig:hr_galex} flag the dwarfs/giants classified in Section 5.1. The \logg\ vs. \Teff\  diagram based on the RAVE/LAMOST stellar parameters roughly agrees with our dwarf/giant classification based on the \emph{Gaia} DR2 HR diagram. The discrepancies should be due to the larger uncertainties of the stellar parameters measured from LAMOST and RAVE spectra. Figure\,\ref{fig:compare_teff_lamost-rave} shows the comparison of spectroscopic \Teff\ (black/red dots for RAVE/LAMOST respectively) with photometric ones presented in Appendix Table\,\ref{tab:tgas-wdfgk-sample}. We can see large deviations between them, $\sim$\,290\,K, especially for high \Teff\ ones (hotter than $\sim$ 6500\,K), which further imply the relatively large uncertainties of stellar parameters from RAVE/LAMOST spectra.

\subsection{High resolution spectroscopic analysis}

At first, we measure the stellar parameters (\Teff, \logg, [Fe/H]) from the high-resolution spectra of 104 WD+AFGK binaries. These were obtained by using the v2019.03.02 version of the freely distributed code iSpec \citep{Blanco2014A&A...569A.111B, Blanco2019MNRAS.486.2075B}. Specifically, we used the spectral synthesis method, utilizing the radiative transfer code SPECTRUM \citep{Gray1994AJ....107..742G}, the MARCS grid of model atmospheres \citep{Gustafsson2008A&A...486..951G}, solar abundances from \citet{Grevesse2007SSRv..130..105G}, and the version 5 of the GES atomic line list \citep{Heiter2015PhyS...90e4010H} between 420 and 920\,nm. Furthermore, before the analysis, we co-added all the duplicated spectra (after applying the RV shift correction) to increase the S/N. 

When performing the fitting, the initial set of atmospheric parameters we used were the photometric \Teff, \logg\ based on SED fitting and the initial [Fe/H] was set to 0.0\,dex. Because of the relatively low S/N achieved during our observations (20--30)\footnote{Note that these values are good enough for measuring reliable RVs.}, we managed to derive the stellar parameters for 55 (which have relatively good spectral quality) of the 104 stars, which are shown in Appendix Table\,\ref{tab:param_high_resolution}. 

\begin{figure}
   \centering
   \includegraphics[width=0.49\textwidth]{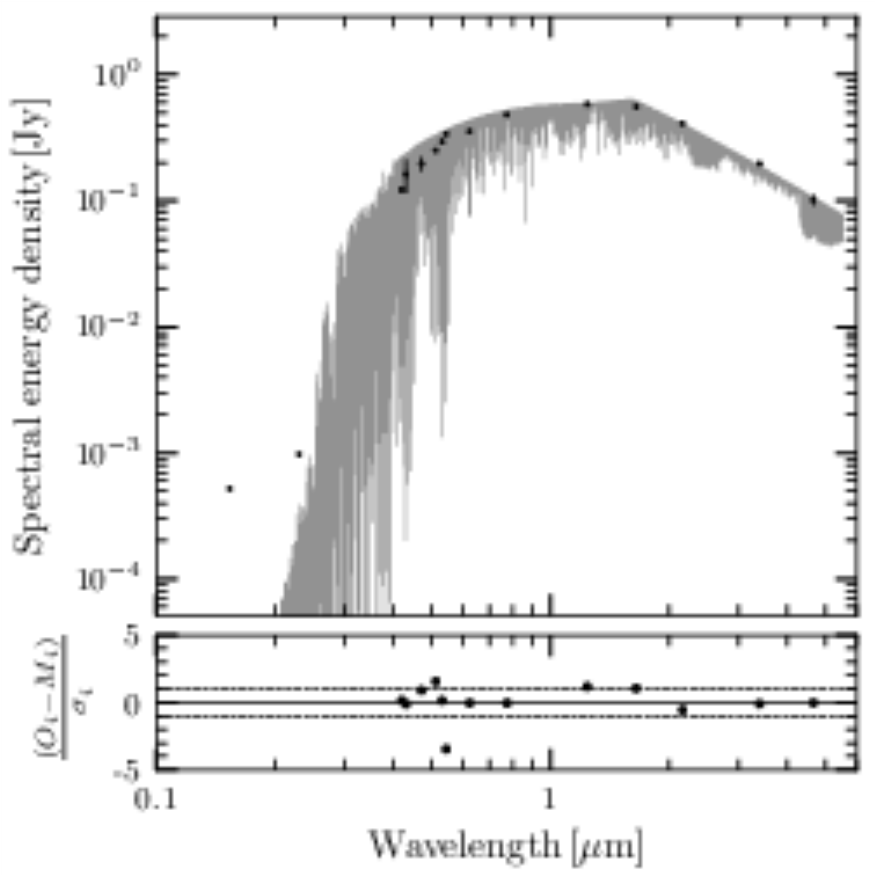} 
   \includegraphics[width=0.49\textwidth]{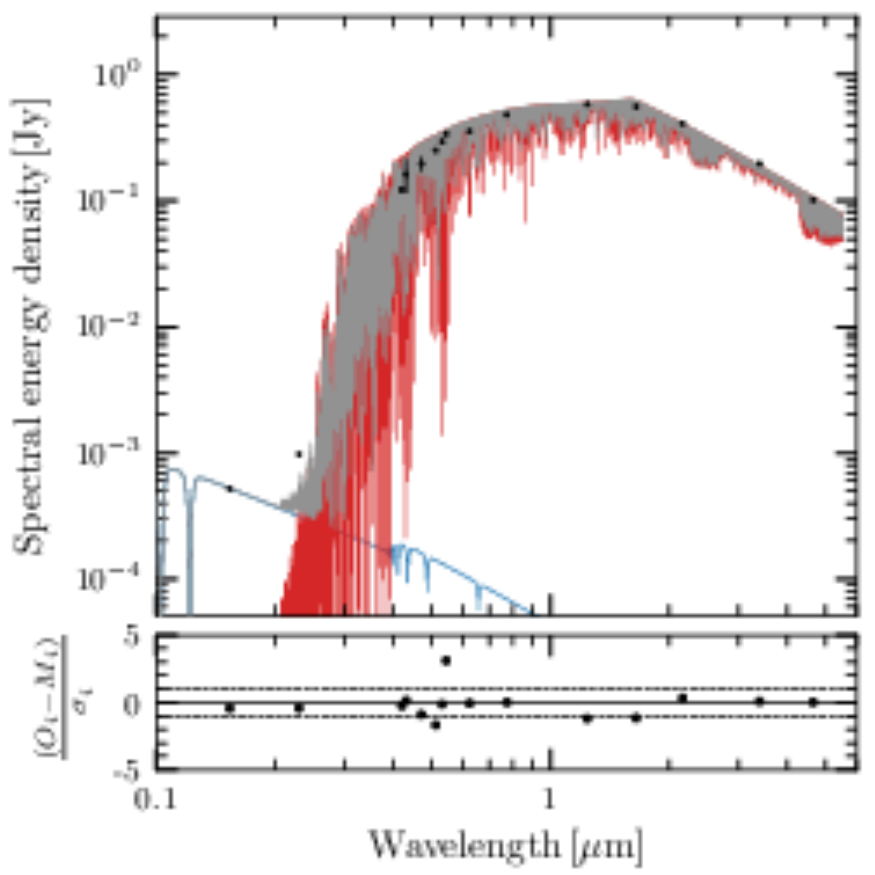} 
   \caption{Top panels: the upper panel shows the single-star SED best-fit for one of the WD+AFGK binary candidates, TYC\,3883-1104-1 (TGAS source ID: 1623107002022578304), while the lower panel shows the error-normalized residuals. Bottom panels: as above, but for the binary-star SED fit. The black dots represent the available photometry for this star. The red and blue spectra represent the best-fitting AFGK and WD models. The composite model spectra of the WD+AFGK binary is shown in gray.}
   \label{fig:sed-double_example}
\end{figure}

Then the stellar masses and radii are derived by using the PARAM 1.3\footnote{http://stev.oapd.inaf.it/cgi-bin/param\_1.3}, which is a Bayesian PARSEC-isochrones \citep{Bressan2012MNRAS.427..127B} fitting code for the estimation of stellar parameters \citep{daSilva2006A&A...458..609D}. The spectroscopic \Teff , \logg\ together with the APASS V\,mag and {\em Gaia} DR2 parallax, are used as the input parameters to estimate the basic stellar parameters. The derived masses and radii are also listed in Appendix Table\,\ref{tab:param_high_resolution}.

\begin{figure}
   \centering
   \includegraphics[width=0.49\textwidth]{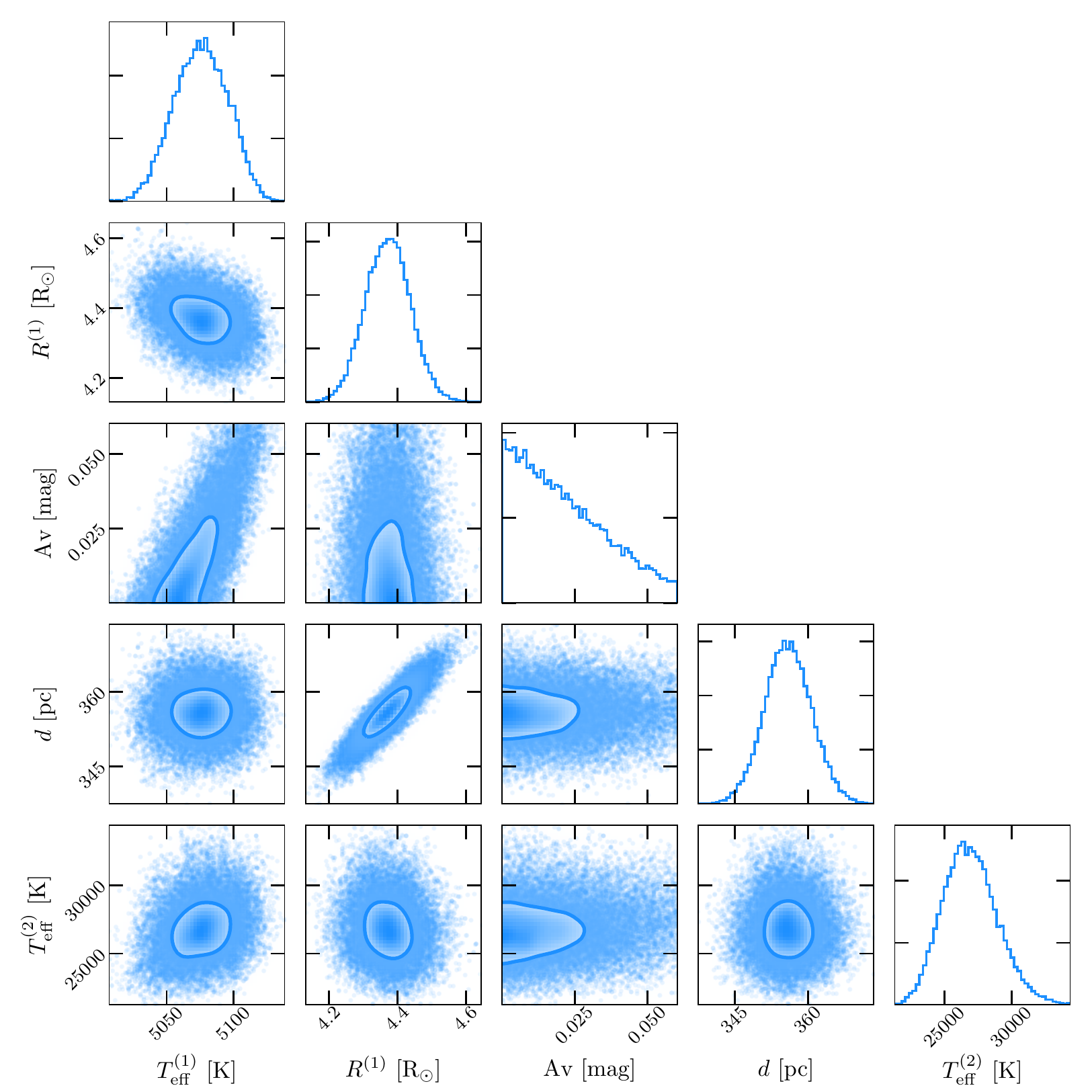} 
   \caption{The corner plots of binary SED fitting solution for TYC\,3883-1101-1, showing the correlations between fitted parameters. Superindex parameter ``1" and ``2" correspond to the AFGK and WD star respectively. The solid blue curves represent the 1$\sigma$ contours.}
   \label{fig:corner_example}
\end{figure}

\begin{figure}
   \centering
   \includegraphics[width=0.49\textwidth]{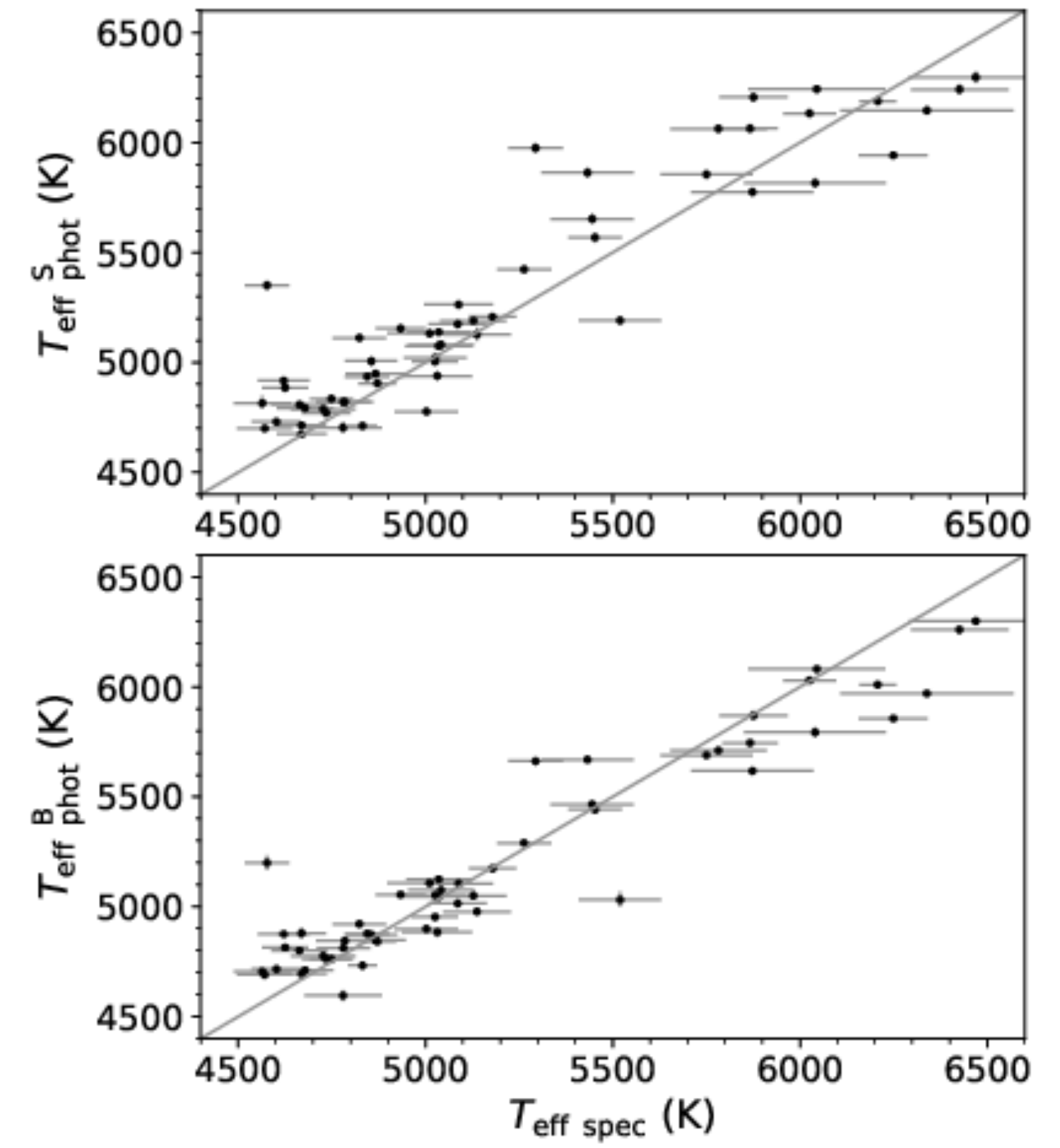} 
   \caption{The comparison of \Teff\ values obtained via the SED fitting, i.e. photometric (upper panel:  \Teff $_\mathrm{phot} ^{S}$ from single SED fitting, bottom panel: \Teff $_\mathrm{phot} ^{B}$ from binary SED fitting) and those determined from the high resolution spectra.}
   \label{fig:compare_teff}
\end{figure}

\subsection{WD contribution to the composite SED}

Following the results of the spectral analysis, we attempted to characterize the WD contribution to the composite SEDs by re-doing the SED-fitting procedure (i.e. obtaining a binary SED-fit solution) and MCMC error analysis outlined in Section 2.1\,--\,2.2; this time, we included the \citet{Koester2010MmSAI..81..921K} grid of synthetic WD spectra with $\log{g} = 8$ (which is representative of the whole WD population) to model the contribution of the UV-bright companions. As external priors, we imposed the high resolution spectroscopic $\log{g} $ and [Fe/H], {\em Gaia} DR2 parallaxes, and the single-star photometric $T_{\rm eff}$, $R$, and $A_{V}$. The composite SED fitting, determines the WD $T_{\rm eff}$ adapting the other parameters. Due to the large deviations of RAVE/LAMOST stellar parameters (as mentioned in Section 5.2), here the binary SED-fitting is only carried out and tested for binary systems with available high resolution spectra.

The bottom panels of Figure\,\ref{fig:sed-double_example} show an example of the binary SED fitting solution for one of our close WD+AFGK candidate (i.e. TYC\,3883-1104-1). For comparison, the top panels show the corresponding single SED fitting solution. We can see that the WD \Teff\ can be estimated after applying the binary SED fitting solution. Figure\,\ref{fig:corner_example} shows the correlations between fitted parameters.

Figure\,\ref{fig:compare_teff} shows the comparison between the high resolution spectroscopic \Teff\ and photometric \Teff\ (upper panel: single SED fitting \Teff $_\mathrm{phot} ^{S}$, bottom panel: binary SED fitting \Teff $_\mathrm{phot} ^{B}$). In the upper panel, the \Teff\ difference has a standard deviation of around 200\,K. While in the bottom panel, when using the binary SED fitting \Teff $_\mathrm{phot} ^{B}$, the \Teff\ difference goes down to 178\,K, which shows a relative improvement of \Teff $_\mathrm{phot}$ when using binary SED fitting. The overall tendency is to find spectroscopic values slightly higher than the photometric ones ($\sim$50--150\,K), a result which was also obtained for single stars by \citet{Zhou2019ApJ...877..104Z}, especially above 5500\,K. Although there is slight disagreement between \Teff $_\mathrm{phot}$ and \Teff $_\mathrm{spec}$ and our target selection is based on \Teff $_\mathrm{phot}$, those targets with available high resolution spectroscopic \Teff\ still well fall within our cuts as shown in Figure\,\ref{fig:colorcut}.

\section{Discussion}

\subsection{Close binary fraction}

In our final sample, we have identified 23 close WD+AFGK binaries (13 dwarfs, 10 giants), and 128 wide binary candidates (64 dwarf, 64 giants) from our RV values. This translates into a close binary fraction of 15\% (dwarf: 17\%, giant: 14\%).

As discussed in \citetalias{Rebassa2017MNRAS.472.4193R}, the RVs measured from higher resolution spectra are more sensitive to detect binaries with longer ($\geq$\,100\,d) orbital periods and lower ($\gtrsim$\,5\,deg) inclinations. Hence the close binary fraction measured from high-resolution spectroscopy should be more reliable. 

If we only take into account the RVs measured from our XL216 and SPM high-resolution spectra, we find 19 close binary system (10 dwarfs, 9 giants) and 85 wide binary candidates (32 dwarfs, 53 giants). Thus the close binary fraction is 18\% (24\% for dwarfs and 15\% for giants). 

The close binary fraction of WD+AFGK binaries harboring dwarf companions is higher than the 10\% fraction we derived in \citetalias{Rebassa2017MNRAS.472.4193R}. This can be explained as follows. First, the results from \citetalias{Rebassa2017MNRAS.472.4193R} are based on a considerably smaller sample than the one used here (63 objects with high-resolution spectra). Second, and more important, the time baseline between the XL216 and SPM observations performed here is considerably larger than in the observations presented in \citetalias{Rebassa2017MNRAS.472.4193R}, which allows to identify longer-period systems. 

Our results also indicate that $\sim$\,14\% of our studied giant AFGK stars display RV variations. It is far away from the scope of this paper to confirm or disprove whether these variations are due to binary membership or pulsations \citep{Wood2004, Nicholls2009} or other intrinsic mechanisms such as solar-type oscillations \citep{Hekker2007}.

\subsection{Stellar parameter distributions of the AFGK stars}

\begin{figure}
   \centering
   \includegraphics[width=0.45\textwidth]{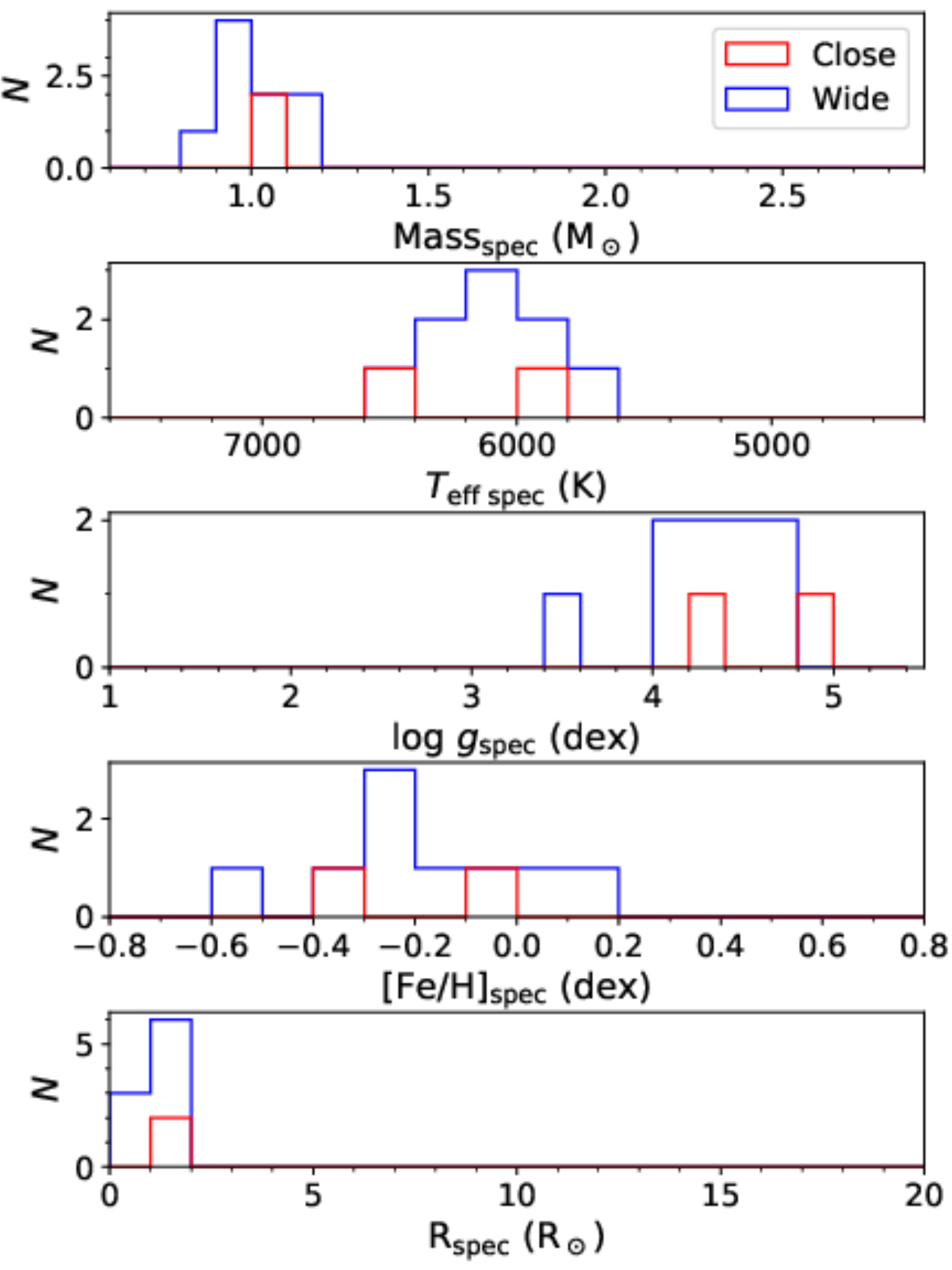} 
   \caption{The histogram distributions of the high resolution spectroscopic parameters of AFGK dwarfs (11) including mass, \Teff, \logg, \feh, and radius. The red and blue lines show the close (2) and wide (9) WD+AFGK candidates respectively. }
   \label{fig:hist_param_spec_dwarfs}
\end{figure}

\begin{figure}
   \centering
   \includegraphics[width=0.45\textwidth]{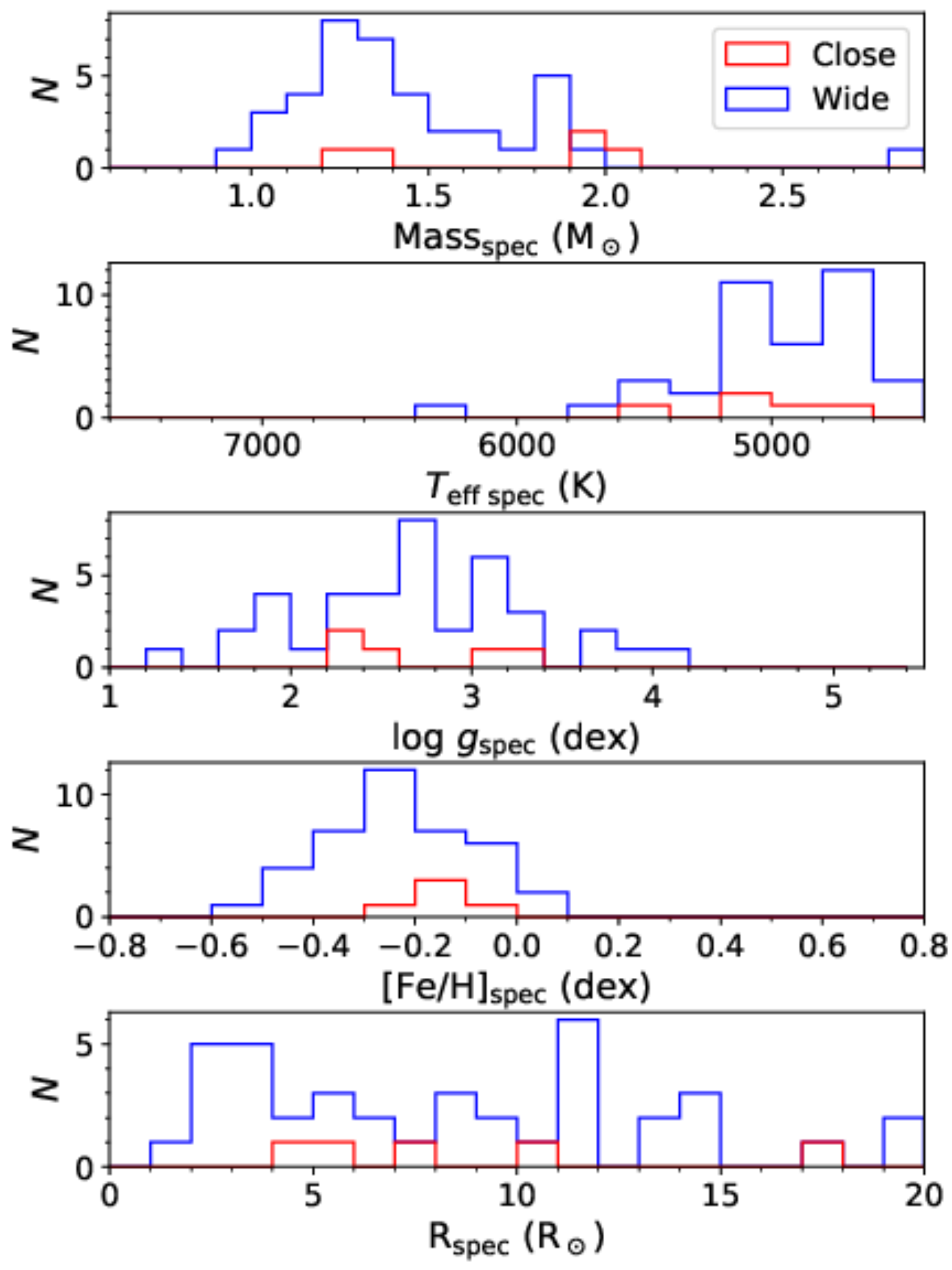} 
   \caption{Similar to Figure\,\ref{fig:hist_param_spec_dwarfs}, but for giants (44). The red and blue lines show the close (5) and wide (39) WD+AFGK candidates respectively.}
   \label{fig:hist_param_spec_giants}
\end{figure}

Based on the high resolution spectroscopic parameters from Table\,\ref{tab:param_high_resolution}, we present the distribution of stellar parameters (i.e. mass, \Teff, \logg, \feh\ and radius) of close/wide systems harboring AFGK dwarfs (11) in Figure\,\ref{fig:hist_param_spec_dwarfs}. The parameter distribution of close/wide (2/9) systems are shown in red/blue color, respectively. From Figure\,\ref{fig:hist_param_spec_dwarfs}, we can see that for these 11 AFGK dwarfs the masses cluster around $\sim$\,1.0\,M$_\sun$. The mass distribution of close and wide binaries are very similar. The same is true for the \Teff, \logg, \feh, and radius distributions. Both the close/wide systems have \Teff\ between 5600--6600\,K, \logg\,$\sim$\,4\,--\,5\,dex, \feh\ between $-$0.6 and $+$0.1\,dex, radius between 0.85\,--\,1.3\,R$_\odot$. A much larger sample of binaries harboring AFGK dwarfs with accurate high resolution spectroscopic parameter determinations is necessary to further investigate possible differences between the stellar parameter distribution of close/wide binaries.

Figure\,\ref{fig:hist_param_spec_giants} is similar to Figure\,\ref{fig:hist_param_spec_dwarfs}, but represents the systems containing giant AFGK companions (44). It becomes obvious that the parameters distribution of these giants are very different from those of arising from the dwarf sample. For the giants, both the close and wide systems have two mass peaks around 1.3 and 1.9\,M$\odot$. The distributions of \Teff, \logg, and radius of close and wide systems in giant samples do not show clear difference neither. Unlike we found in the dwarf samples, the \Teff\ of the giant samples is clustered around a relatively colder temperature of $\sim$\,5000\,K. The \feh\ distribution of the giant samples are between $-$0.6 and  0.1\,dex, which are very similar to the dwarf samples. It is also worth noting that the wide systems in the giant sample seem to be slightly metal poor (peaks around $-$0.3\ to $-$0.2\,dex) as compared to those that are part of close systems (with a peak between $-$0.2 and $-$0.1\,dex), which maybe due to the observational selection effect.

\begin{figure}[htbp] 
   \centering
   \includegraphics[width=0.45\textwidth]{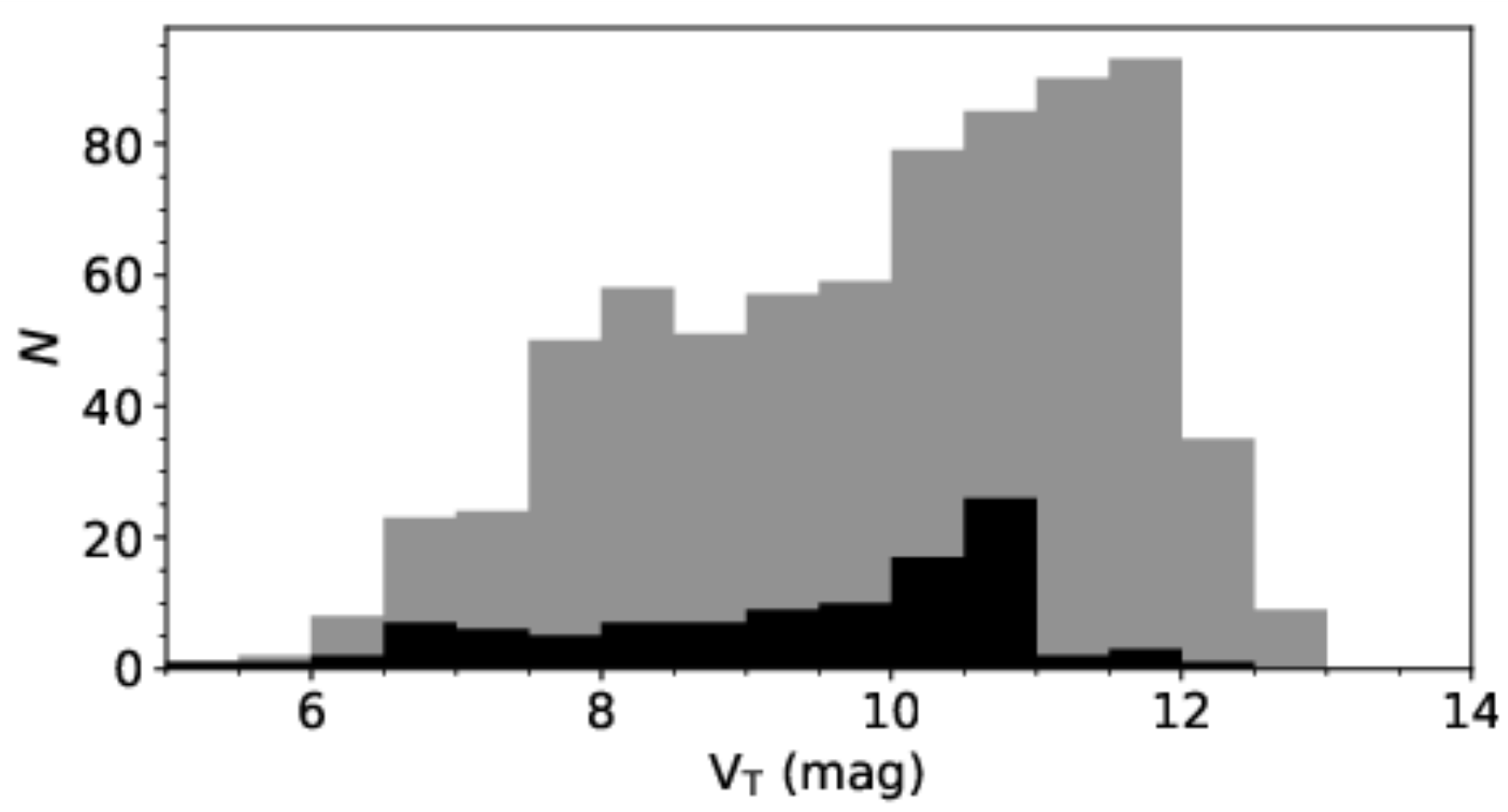} 
   \caption{The histogram distribution of V$_\mathrm{T}$ magnitudes of all the WD+AFGK candidates (gray filled steps) and high resolution spectroscopic observed ones by XL216 and SPM (black filled steps).}
   \label{fig:hist_VT}
\end{figure}

Furthermore, the stellar parameters distributions of close/wide systems we obtained here may suffer from observational selection effects. The telescopes we used are of two-meters (XL216 and SPM), which can only observe the very bright stars when equipped with a high resolution Echelle spectrograph. In order to observe as many objects as possible and to improve the observation efficiency, we prioritized observations of the brighter objects in our sample. To put into contest this effect, in Figure\,\ref{fig:hist_VT} we show the histogram distributions of the V$_\mathrm{T}$ magnitudes of all the WD+AFGK candidates compared to those we observed at high-resolution. We can clearly see that the targets we observed are generally brighter than 11\,mag, which is close to the limiting magnitude of two-meters telescope equipped with high resolution spectrograph.

\begin{figure}
   \centering
   \includegraphics[width=0.45\textwidth]{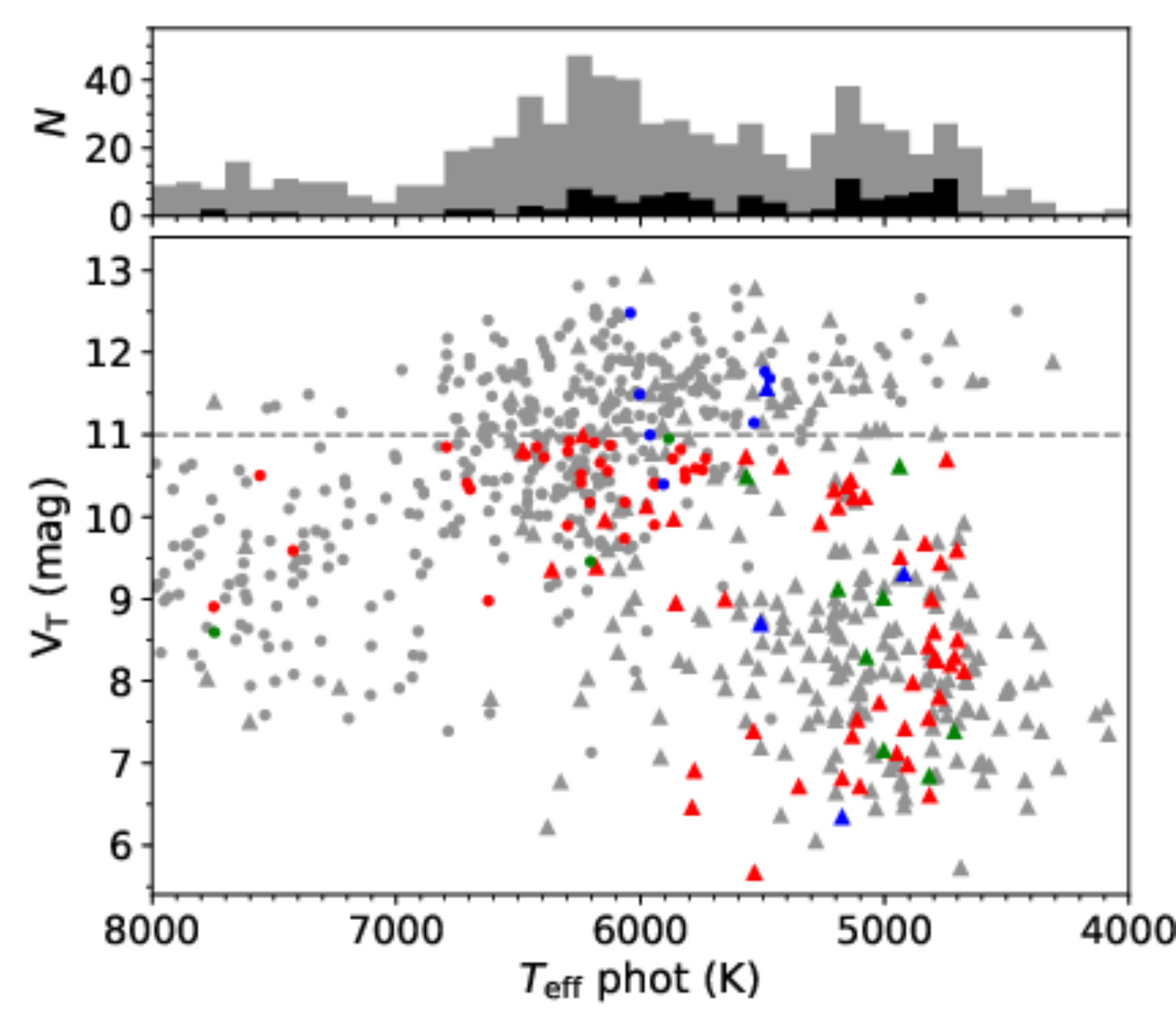} 
   \caption{The V$_\mathrm{T}$ magnitude vs. photometric \Teff\ for all the WD+AFGK candidates (gray dots), and those observed by XL216 (red dots), SPM (blue dots), XL216+SPM (i.e. both observed by XL216 and SPM, green dots). The dots and triangles mark the dwarfs/giants respectively. The horizontal gray dashed line marks the V$_\mathrm{T}$\,=\,11\,mag line. The upper small panel shows the histogram distribution of photometric \Teff\ of those high resolution observed ones (black filled steps) and all the WD+AFGK candidates (gray filled steps). The horizontal gray dashed lines in the bottom panel shows the V$_\mathrm{T}$\,=\,11\,mag. }
   \label{fig:V_Teff_obs}
\end{figure}

To investigate possible observational selection effects on the parameter distributions, the bottom panel of Figure\,\ref{fig:V_Teff_obs} shows the V$_\mathrm{T}$ vs. photometric \Teff\ for all the WD+AFGK candidates and those with high resolution observations. The upper panel shows their corresponding histogram distribution of \Teff. Inspection of the figure reveals the fraction of observed targets are relatively low near 6000\,K (most are dwarfs), while most of the targets near 5000\,K were observed due to their intrinsic brightness (most are giants). Thus, we can easily explain the difference between the  \Teff\ distribution in Figure\,\ref{fig:hist_param_spec_dwarfs} and Figure\,\ref{fig:hist_param_spec_giants}. But this observational selection effects do not affect the close binary fractions we measured in Section\,6.1.

\subsection{Stellar parameter distributions of WD}

\begin{figure}
   \centering
   \includegraphics[width=0.45\textwidth]{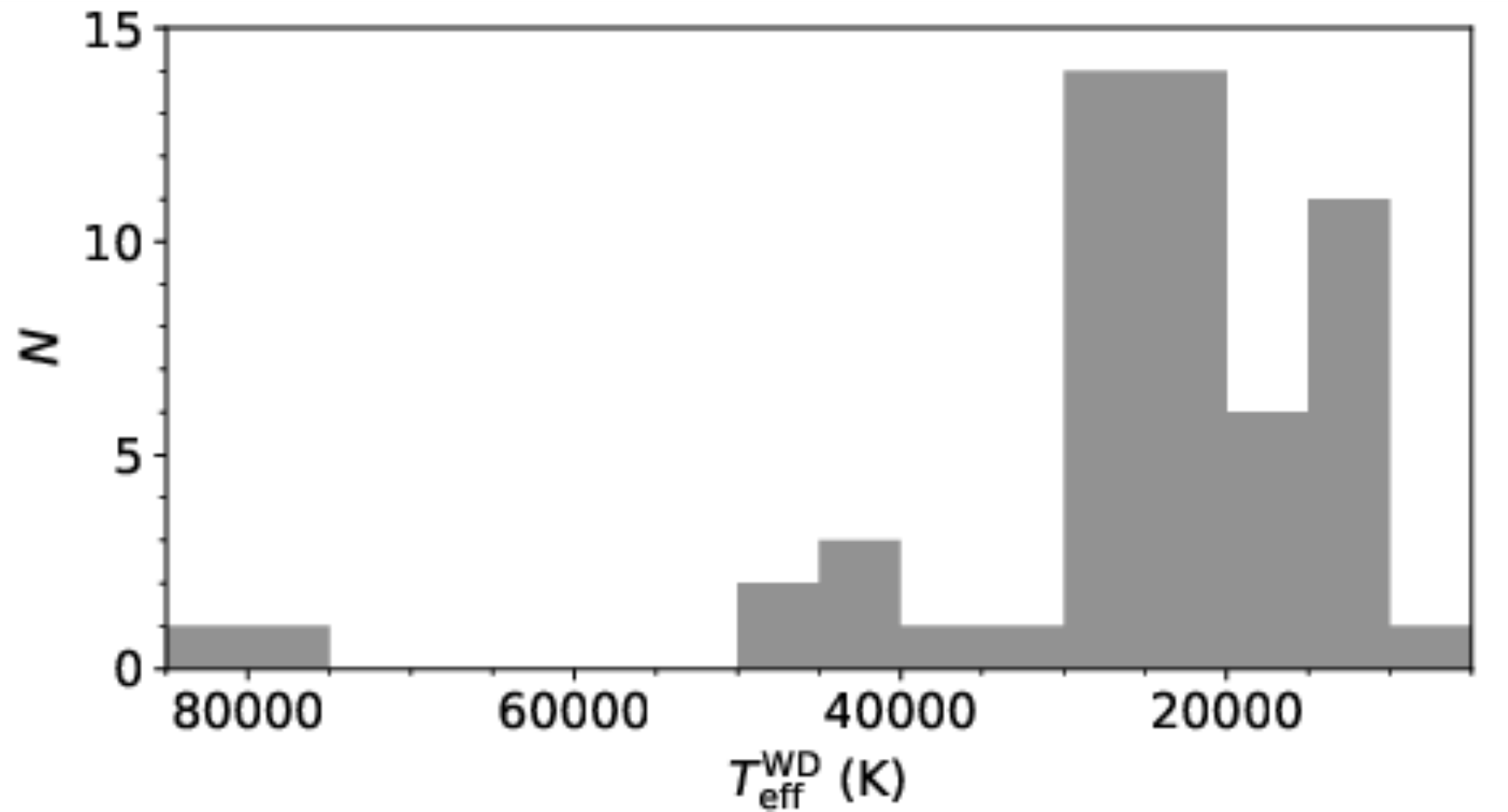} 
   \caption{The histogram distributions of $T_\mathrm{eff}^\mathrm{WD}$ determined from binary SED fitting. }
   \label{fig:hist_WDparam}
\end{figure}

By using the binary SED fitting solutions for those WD+AFGK candidates with available high resolution spectroscopic stellar parameters (in Appendix Table\,\ref{tab:param_high_resolution}), the WD \Teff\ can be determined at the same time as described in Section 5.4. Figure\,\ref{fig:hist_WDparam} plots the histogram of the $T_\mathrm{eff}^\mathrm{WD}$. We can see that the distribution of WD \Teff\ has a peak around 10\,000\,--\,30\,000\,K, which agrees with the $T_\mathrm{eff}^\mathrm{WD}$ distributions from the WD+M binary sample \citep{Rebassa2016MNRAS.458.3808R}.

\subsection{Space density of WD+AFGK binaries}

\begin{figure}
   \centering
   \includegraphics[width=0.45\textwidth]{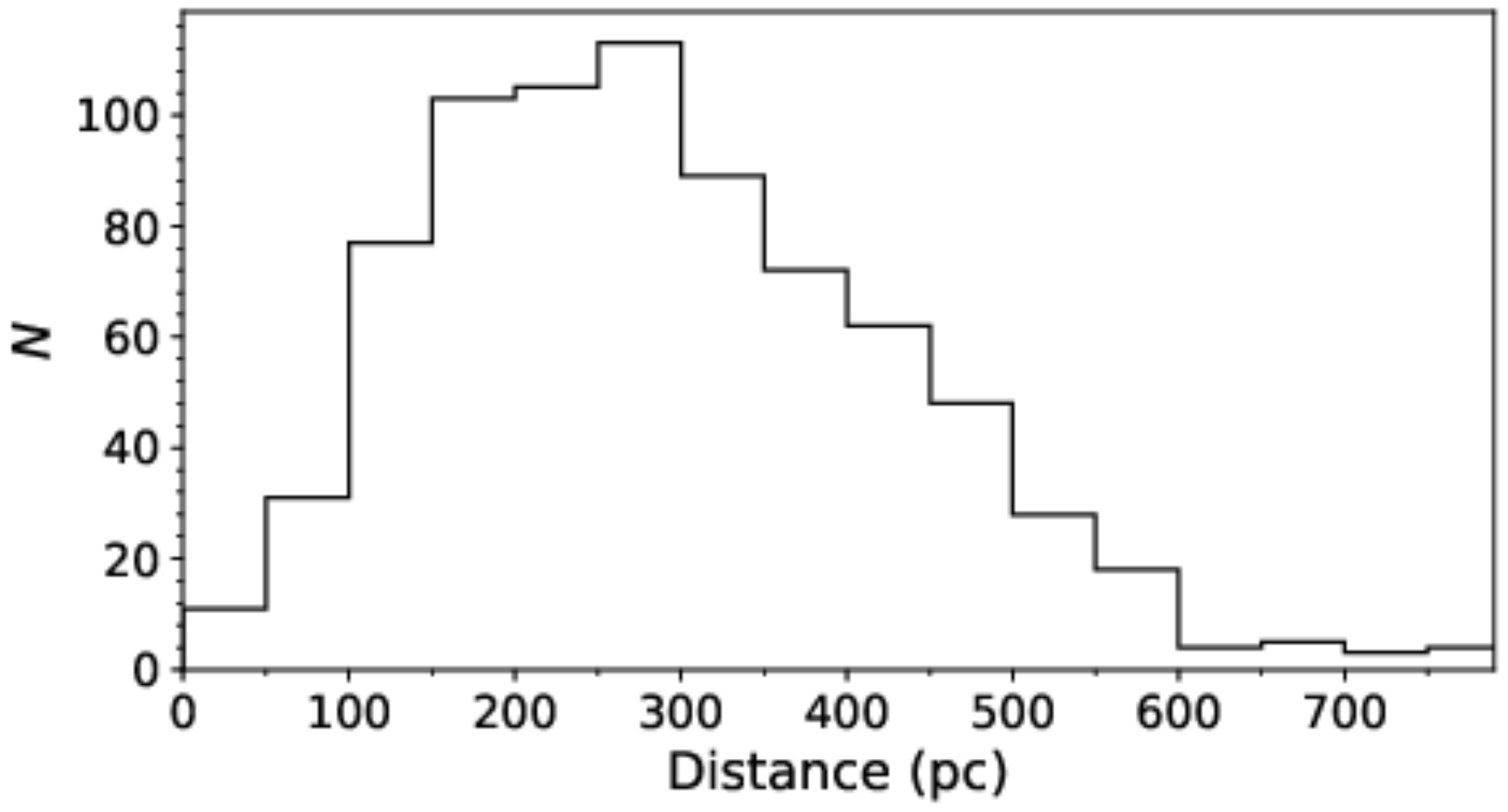} 
   \caption{The histogram distributions of distances of the WD+AFGK binaries.}
   \label{fig:hist_distance}
\end{figure}

Our WD+AFGK binary sample has available $Gaia$ parallaxes, hence it is possible to infer their distances. We obtain these values from the catalog of \citet{Bailer2018AJ....156...58B} and the corresponding histogram is shown in Figure\,\ref{fig:hist_distance}, where we can see that our binaries are located in the 10\,$\sim$\,858\,pc range, peaking at $\sim$\,250\,--\,300\,pc. For comparison, \citet{Toonen2017A&A...602A..16T} present a 20\,pc sample of WD plus AFGKM main-sequence binaries (WDMS), which is more complete and includes 2 unresolved WDMS and 24 resolved WDMS. Our WD+AFGK sample extends this census up to $\sim$\,850\,pc, although it is not complete due to the possible incompleteness of \emph{Gaia} or \emph{GALEX}, especially in the Galactic plane, which isn't covered by \emph{GALEX}.

\citet{Jimenez2018MNRAS.480.4505J} analyses the completeness of the WD population accessible by \emph{Gaia} as a function of the parallax relative error (see their Figure 1), from which it is possible to see how the distance affects the completeness of the \emph{Gaia} WD sample. If we assume a similar distance effect in our binary sample, and considering that the distance distribution of our WD+AFGK binaries peaks around 300\,pc, we can assume a completeness of $\sim$\,50\% for our sample. If we further add the incompleteness of \emph{GALEX} and the fact that we are biased towards the detection of relatively hot WDs, then the incompleteness of our WD+AFGK sample should be severely lower than 50\%.

From the distance information, we can estimate the space density $\rho$ of our WD+AFGK binaries just integrating the number of objects in the volume considered and incorporating the scale height of the thin disk \citep[322\,pc,][]{Chen2017MNRAS.464.2545C} as a weighting factor in the integral \citep{Schreiber2003A&A...406..305S}. The calculation yields $\rho$\,=\,1.9$\times10^{-6}$\,pc$^{-3}$. However, as our sample is not complete, this value should be considered as a lower limit. Furthermore, with the aim of considering a volume-limited sample rather than a magnitude-limited one, we also estimate the space density for distances within 200\,pc, which results in 10.5$\times10^{-6}$\,pc$^{-3}$. Given that the WDs in these WD+AFGK binaries are generally hot, henceforth detectable via their UV excess, this result should also be considered as a lower limit.

\section{Summary}

As one of a series of papers aiming at constraining the past and future evolution of close compact binaries, here we presented the identification of WD+AFGK binaries from the TGAS and \emph{Gaia} DR2 databases. We selected 814 WD+AFGK binary candidates through the detection of UV-excess, out of which we selected 775 candidates after excluding possible contaminants. An extensive high resolution spectroscopic follow-up campaign has been carried out to obtain at least two high precision RVs separated at different nights for each of the selected WD+AFGK binaries. 214 high resolution spectra (R\,$\sim$\,49\,800) were obtained from the Xinglong 2.16\,m telescope for 93 WD+AFGK binary, and 50 high resolution (R\,$\sim$ 20\,000) spectra were obtained for 22 WD+AFGK binaries. Furthermore, all the available spectroscopic sky survey data like LAMOST DR6 low resolution spectra and RAVE DR5 medium resolution spectra were also used to identify as many close binaries as possible. 

We provide 517 RVs for 275 of our WD+AFGK binaries, from which we identify 23 close binaries via RV variations and 128 likely wide binaries. Interestingly, we find a relatively large percentage of WD+AFGK systems containing giants displaying RV variations. The close binary fraction we derive for WD+AFGK containing dwarf stars is around 24\%. The close binary fraction for dwarf companions (24\%) is higher than that for giant companions (15\%). The atmospheric parameters (\Teff, \logg, [Fe/H]), as well as stellar mass and radius of the AFGK companions are provided from the high resolution spectroscopy. The stellar parameters and mass distributions of the close and wide binaries are similar. Based on the Gaia distance, a lower limit of space density of WD+AFGK binary candidates is estimated to be 1.9$\times10^{-6}$\,pc$^{-3}$, which increases to 10.5$\times10^{-6}$\,pc$^{-3}$ for samples within 200\,pc.

Most of the close binaries found from this work are intrinsically bright (most are brighter than 11\,mag), thus are easy to be followed-up in the near future for measuring their orbital periods and component masses. This will allow us to study the past and future evolution of these systems and thus improve our understanding of common envelope evolution  and investigate possible formation channels for SN Ia.

\section*{Acknowledgments}

This work is partially supported by National Natural Science Foundation of China 11903048, 11833006, U1831209, and 11873057. RR has received funding from the postdoctoral fellowship programme Beatriu de Pin\'os, funded by the Secretary of Universities and Research (Government of Catalonia) and by the Horizon 2020 programme of research and innovation of the European Union under the Maria Sk\l{}odowska-Curie grant agreement No 801370. ARM acknowledges support from the MINECO under the Ram\'on  y Cajal programme (RYC-2016-20254) and the AYA2017-86274-P grant, and the AGAUR grant SGR-661/2017. MSH acknowledges support through a Fellowship for National PhD students from ANID, grant number 21170070. SGP acknowledges the support of the STFC Ernest Rutherford Fellowship. MRS thanks for support from FONDECYT (grant number 1181404).

Based on observations performed at Xinglong 2.16\,m telescope and the 2.12\,m telescope in San Pedro M\'artir Observatory. We acknowledge the support of the staff of the Xinglong 2.16m telescope. This work was partially supported by the Open Project Program of the Key Laboratory of Optical Astronomy, National Astronomical Observatories, Chinese Academy of Sciences.

This work has made use of data from the European Space Agency (ESA) mission Gaia (https://www.cosmos.esa.int/gaia), processed by the Gaia Data Processing and Analysis Consortium (DPAC, https://www.cosmos.esa.int/web/gaia/dpac/consortium). Funding for the DPAC has been provided by national institutions, in particular the institutions participating in the Gaia Multilateral Agreement.

This work has made use of data products from the Guoshoujing Telescope (the Large Sky Area Multi-Object Fibre Spectroscopic Telescope, LAMOST). LAMOST is a National Major Scientific Project built by the Chinese Academy of Sciences. Funding for the project has been provided by the National Development and Reform Commission. LAMOST is operated and managed by the National Astronomical Observatories, Chinese Academy of Sciences. Funding for RAVE (www.rave-survey.org) has been provided by institutions of the RAVE participants and by their national funding agencies. This research has made use of the SIMBAD database, operated at CDS, Strasbourg, France.

\software{SciPy \citep{Virtanen2020NatMe..17..261V}, Astropy \citep{Astropy2013A&A...558A..33A, Astropy2018AJ....156..123A}, LMFIT \citep{newville_matthew_2014_11813}, iSpec \citep[v2019.03.02;][]{Blanco2014A&A...569A.111B,Blanco2019MNRAS.486.2075B}, SPECTRUM \citep{Gray1994AJ....107..742G}, MARCS \citep{Gustafsson2008A&A...486..951G}, IRAF \citep{Tody1986SPIE..627..733T,Tody1993ASPC...52..173T}, emcee \citep{Foreman2019JOSS....4.1864F}, PARAM \citep[v1.3;][]{daSilva2006A&A...458..609D}}

\setcounter{table}{0}
\renewcommand{\thetable}{A\arabic{table}}

\begin{longrotatetable}
\setlength{\tabcolsep}{2.5pt}
\begin{splitdeluxetable*}{lcrrrrrrrrBrrrrrrrrrccccc} 
\tablecaption{The 814 selected WD+AFGK binary candidates\label{tab:tgas-wdfgk-sample}.}
\tabletypesize{\scriptsize}
\tablewidth{0pt} 
\tablehead{ 
\colhead{Name} & \colhead{Source ID (TGAS)} & \colhead{$\varpi$ (TGAS)} & \colhead{B$_\mathrm{T}$}  & \colhead{V$_\mathrm{T}$} & \colhead{FUV}  & \colhead{NUV} & 
\colhead{Source ID (DR2)} & \colhead{$\alpha$ (DR2)} & \colhead{$\delta$ (DR2)} &\colhead{$\varpi$ (DR2)} &\colhead{$\mu_\alpha$} & \colhead{$\mu_\delta$} &\colhead{$G$} &\colhead{$G_\mathrm{BP}$} & \colhead{$G_\mathrm{RP}$} &
\colhead{\Teff} & \colhead{R} & \colhead{$A_V$} &  
\colhead{Dwarf} & \colhead{SIMBAD Classification} & \colhead{Contaminant} & \colhead{Spec} & \colhead{Flag} \\ 
\colhead{} & \colhead{} & \colhead{(mas)} & \colhead{(mag)} & \colhead{(mag)} & \colhead{(mag)} & \colhead{(mag)} &
\colhead{} & \colhead{($^\circ$)} & \colhead{($^\circ$)} & \colhead{(mas)} & \colhead{(mas yr$^{-1}$)} & \colhead{(mas yr$^{-1}$)}  & \colhead{(mag)} & \colhead{(mag)} & \colhead{(mag)} &
\colhead{(K)}  & \colhead{(R$_\odot$)} & \colhead{(mag)} &
\colhead{}  & \colhead{} & \colhead{} & \colhead{} & \colhead{} 
} 
\startdata \\
TYC\,1006-4-1    & 4500804497016311040 & 2.847$\pm$0.298 &  9.214$\pm$0.017 &  7.986$\pm$0.011 & 17.672$\pm$0.057 & 15.751$\pm$0.017 & 4500804501313209344 & 266.9840554690 & 14.7535113036 & 2.313$\pm$0.052 &     0.092$\pm$0.075 & $-$18.014$\pm$0.090 &  7.564 &  8.132 &  6.892 & 4885.0$_{-20.0} ^{+20.0}$ & 18.86$_{-0.40} ^{+0.33}$ & 0.25$_{-0.02} ^{+0.02}$ & N & Star & N & $-$ & $-$ \\
TYC\,1007-942-1  & 4476095172207240704 & 3.592$\pm$0.465 &  9.687$\pm$0.024 &  9.436$\pm$0.022 & 13.192$\pm$0.007 & 13.485$\pm$0.003 & 4476095176502659584 & 269.8479409744 &  7.7092128037 & 3.724$\pm$0.043 &  $-$6.981$\pm$0.068 &  $-$7.211$\pm$0.065 &  9.391 &  9.512 &  9.204 & 7625.0$_{-35.0} ^{+37.0}$ &  1.81$_{-0.02} ^{+0.02}$ & 0.08$_{-0.01} ^{+0.01}$ & Y & Star & N & $-$ & $-$ \\
TYC\,1010-2134-1 & 4478719603379517568 & 2.449$\pm$0.265 & 10.324$\pm$0.027 & 10.098$\pm$0.029 & 14.372$\pm$0.013 & 13.674$\pm$0.004 & 4478719607687083904 & 275.6919592219 &  8.8715466865 & 2.261$\pm$0.052 &  $-$3.762$\pm$0.074 &     5.089$\pm$0.083 &  9.956 & 10.102 &  9.734 & 7438.0$_{-32.0} ^{+32.0}$ &  2.41$_{-0.05} ^{+0.05}$ & 0.10$_{-0.01} ^{+0.01}$ & Y & Star & N & $-$ & $-$ \\
TYC\,1010-403-1  & 4481708041623948672 & 2.127$\pm$0.285 & 11.400$\pm$0.063 & 11.412$\pm$0.098 & 14.684$\pm$0.011 & 14.173$\pm$0.004 & 4481708045927762944 & 274.5910736665 &  9.1088137354 & 1.727$\pm$0.044 &     4.205$\pm$0.073 &     2.734$\pm$0.078 & 11.056 &    $-$ &    $-$ & 7745.0$_{-66.0} ^{+63.0}$ &  1.94$_{-0.04} ^{+0.04}$ & 0.30$_{-0.02} ^{+0.02}$ & N & Star & N & LAMOST & $-$ \\
TYC\,1012-788-1  & 4494712721561721728 & 2.733$\pm$0.266 &  9.469$\pm$0.020 &  8.540$\pm$0.014 & 14.404$\pm$0.015 & 13.492$\pm$0.006 & 4494712725861730176 & 270.0509389652 & 10.5163424463 & 2.060$\pm$0.067 &     2.853$\pm$0.119 &     2.763$\pm$0.119 &  8.184 &  8.682 &  7.556 & 5356.0$_{-28.0} ^{+24.0}$ & 13.55$_{-0.37} ^{+0.31}$ & 0.47$_{-0.02} ^{+0.01}$ & N & Star & N & $-$ & $-$ \\
TYC\,1020-875-1  & 4498214219417652608 & 1.930$\pm$0.247 & 12.596$\pm$0.193 & 12.183$\pm$0.170 & 18.254$\pm$0.083 & 15.789$\pm$0.018 & 4498214219417652608 & 271.7589501457 & 14.6368243940 & 1.836$\pm$0.041 &     5.025$\pm$0.074 &     3.240$\pm$0.073 & 11.923 & 12.172 & 11.533 & 6595.0$_{-31.0} ^{+32.0}$ &  1.56$_{-0.02} ^{+0.03}$ & 0.14$_{-0.01} ^{+0.01}$ & Y & Star & N & LAMOST & $-$ \\
TYC\,1023-2378-1 & 4477623459012735488 & 4.301$\pm$0.306 & 13.069$\pm$0.262 & 11.763$\pm$0.134 & 17.051$\pm$0.047 & 16.623$\pm$0.027 & 4477623459012735488 & 277.2176435238 &  7.8490471591 & 4.253$\pm$0.048 &     6.571$\pm$0.083 &  $-$1.306$\pm$0.088 & 11.645 & 12.024 & 11.120 & 5895.0$_{-28.0} ^{+28.0}$ &  1.00$_{-0.00} ^{+0.00}$ & 0.26$_{-0.02} ^{+0.02}$ & Y & Star & N & $-$ & $-$ \\
TYC\,1027-1804-1 & 4480505210263029632 & 3.045$\pm$0.276 & 12.458$\pm$0.187 & 11.551$\pm$0.110 & 18.275$\pm$0.087 & 16.689$\pm$0.030 & 4480505214573963520 & 276.9209151405 & 10.5316866686 & 3.917$\pm$0.052 &  $-$4.483$\pm$0.085 &     0.261$\pm$0.087 & 11.290 & 11.659 & 10.765 & 5718.0$_{-19.0} ^{+19.0}$ &  1.29$_{-0.01} ^{+0.01}$ & 0.07$_{-0.01} ^{+0.01}$ & Y & Star & N & $-$ & $-$ \\
TYC\,1031-707-1  & 4485165902619270656 & 3.399$\pm$0.334 &  9.415$\pm$0.020 &  9.198$\pm$0.018 & 12.932$\pm$0.008 & 12.859$\pm$0.005 & 4485165906920932864 & 276.4871821339 & 12.7786034894 & 3.691$\pm$0.041 &  $-$8.628$\pm$0.070 &  $-$5.475$\pm$0.072 &  9.142 &  9.248 &  8.986 & 7861.0$_{-37.0} ^{+36.0}$ &  1.89$_{-0.02} ^{+0.02}$ & 0.00$_{-0.00} ^{+0.00}$ & Y & Variable Star & N & $-$ & $-$ \\
TYC\,103-810-1   & 3234843885583146880 & 2.494$\pm$0.305 & 10.170$\pm$0.033 &  9.972$\pm$0.038 & 13.259$\pm$0.006 & 13.284$\pm$0.006 & 3234843889879946880 &  77.2968945732 &  2.1784489294 & 2.281$\pm$0.043 &  $-$2.006$\pm$0.081 & $-$12.144$\pm$0.054 &  9.881 & 10.019 &  9.650 & 7101.0$_{-30.0} ^{+28.0}$ &  2.95$_{-0.02} ^{+0.03}$ & 0.18$_{-0.01} ^{+0.00}$ & Y & Star & N & $-$ & $-$ \\
TYC\,110-755-1   & 3240025986963617792 & 7.477$\pm$0.229 & 11.202$\pm$0.074 & 10.570$\pm$0.063 & 16.882$\pm$0.039 & 14.936$\pm$0.009 & 3240025986963617792 &  76.7235053441 &  6.1154965611 & 7.326$\pm$0.065 & $-$13.519$\pm$0.093 &  $-$9.512$\pm$0.076 & 10.300 & 10.663 &  9.808 & 5746.0$_{-23.0} ^{+22.0}$ &  1.12$_{-0.01} ^{+0.01}$ & 0.15$_{-0.01} ^{+0.01}$ & Y & Star & N & $-$ & $-$ \\
TYC\,1117-2238-1 & 1758854975332493568 & 3.363$\pm$0.280 &  9.554$\pm$0.024 &  8.896$\pm$0.018 & 17.615$\pm$0.057 & 13.374$\pm$0.005 & 1758854979627147776 & 318.5411607094 & 13.2453330926 & 3.099$\pm$0.039 &  $-$2.073$\pm$0.072 &  $-$7.297$\pm$0.061 &  8.635 &  8.970 &  8.170 & 6049.0$_{-23.0} ^{+23.0}$ &  5.12$_{-0.07} ^{+0.07}$ & 0.21$_{-0.01} ^{+0.01}$ & N & Star & N & $-$ & $-$ \\
TYC\,1134-190-1  & 1768992369459334272 & 3.283$\pm$0.248 &  9.789$\pm$0.022 &  9.209$\pm$0.017 & 18.146$\pm$0.046 & 14.134$\pm$0.004 & 1768992369459334272 & 328.6807253391 & 14.5576497717 & 3.329$\pm$0.048 &  $-$8.680$\pm$0.072 & $-$22.244$\pm$0.069 &  8.697 &  9.064 &  8.196 & 5517.0$_{-18.0} ^{+18.0}$ &  5.13$_{-0.08} ^{+0.08}$ & 0.00$_{-0.00} ^{+0.00}$ & N & Eclipsing binary of Algol type & Y & $-$ & $-$ \\
TYC\,1134-414-1  & 1768986562663551104 & 2.776$\pm$0.278 &  9.512$\pm$0.019 &  9.243$\pm$0.017 & 14.931$\pm$0.010 & 13.853$\pm$0.004 & 1768986566959444096 & 328.6396538629 & 14.5347921319 & 2.873$\pm$0.058 &    14.001$\pm$0.084 &  $-$1.838$\pm$0.089 &  9.178 &  9.308 &  8.980 & 7538.0$_{-30.0} ^{+30.0}$ &  2.63$_{-0.05} ^{+0.05}$ & 0.08$_{-0.01} ^{+0.01}$ & Y & Eclipsing binary & Y & $-$ & $-$ \\
\nodata & \nodata & \nodata & \nodata & \nodata & \nodata & \nodata & \nodata & \nodata & \nodata & \nodata & \nodata & \nodata & \nodata & \nodata & \nodata & \nodata & \nodata & \nodata & \nodata & \nodata & \nodata & \nodata & \nodata \\
\enddata   
\tablecomments{Here we  list the  Name, TGAS source ID, parallax, Tycho $B_\mathrm{T}V_\mathrm{T}$ magnitudes, \emph{GALEX} photometry, Gaia DR2 coordinate on epoch of 2015.5, parallax, proper motion, $G$/$G_\mathrm{BP}$/$G_\mathrm{RP}$ magnitudes, the SED fitting results, the dwarf/giant classification (``Dwarf=Y" shows dwarf, ``N" is giant), the SIMBAD classification, the contaminant flag (``Contaminant=Y" shows the possible contaminant, ``N" shows our final sample), and whether it has available spectra from RAVE DR5 or LAMOST DR6 (column ``Spec"). The last column marks if it has already been published before, where ``flag=a" means published in the RAVE WD+AFGK sample from \citetalias{Parsons2016MNRAS.463.2125P}, ``flag=b" shows those published in the LAMOST WD+AFGK sample from \citetalias{Rebassa2017MNRAS.472.4193R}, and ``flag=$-$" shows the new ones which are unpublished before. The entire table is provided in the electronic version of the paper.}
\end{splitdeluxetable*}
\end{longrotatetable}

\begin{deluxetable*}{lrrrrrr}
\centering
\setlength{\tabcolsep}{3.0pt}
\tablecaption{The spectroscopic information of TGAS-RAVE/LAMOST WD+AFGK binaries\label{tab:param_tgas-rave-lamost}}
\tabletypesize{\scriptsize}
\tablewidth{0pt} 
\tablehead{ 
\colhead{Name} & \colhead{HJD} & \colhead{S/N} & \colhead{\Teff} & \colhead{\logg} &\colhead{Metallicity} & \colhead{Flag} \\
\colhead{} & \colhead{(d)} & \colhead{} & \colhead{(K)} & \colhead{(dex)} & \colhead{(dex)} & \colhead{} 
} 
\startdata
TYC\,1010-403-1   & 2457860.34272 & 220.67 & \nodata           & \nodata         & \nodata            & LAMOST \\
TYC\,1010-403-1   & 2457917.19322 & 223.53 & \nodata           & \nodata         & \nodata            & LAMOST \\
TYC\,1020-875-1   & 2457528.30082 &  87.18 & 6808.76$\pm$23.13 & 4.053$\pm$0.038 & $-$0.139$\pm$0.022 & LAMOST \\
TYC\,12-20-1      & 2456199.24772 &   1.56 & \nodata           & \nodata         & \nodata            & LAMOST \\
TYC\,1246-582-1   & 2457018.06924 & 261.65 & \nodata           & \nodata         & \nodata            & LAMOST \\
TYC\,1246-582-1   & 2457662.37084 &   5.43 & \nodata           & \nodata         & \nodata            & LAMOST \\
TYC\,1246-850-1   & 2457662.37087 & 493.05 & \nodata           & \nodata         & \nodata            & LAMOST \\
TYC\,1287-1768-1  & 2456946.36737 & 439.64 & \nodata           & \nodata         & \nodata            & LAMOST \\
TYC\,1380-957-1   & 2456283.23496 & 156.63 & 5915.45$\pm$19.50 & 4.321$\pm$0.032 &    0.130$\pm$0.017 & LAMOST \\
TYC\,1389-1680-1  & 2458168.07737 & 205.37 & 5719.85$\pm$21.30 & 3.987$\pm$0.034 &    0.008$\pm$0.018 & LAMOST \\
TYC\,1428-81-1    & 2457444.25195 &   6.34 & \nodata           & \nodata         & \nodata            & LAMOST \\
TYC\,1451-111-1   & 2456021.14327 & 142.35 & \nodata           & \nodata         & \nodata            & LAMOST \\
TYC\,1451-111-1   & 2457435.39474 & 267.49 & 6217.27$\pm$11.67 & 4.255$\pm$0.017 & $-$0.198$\pm$0.009 & LAMOST \\
TYC\,1451-111-1   & 2457438.31629 &   3.65 & \nodata           & \nodata         & \nodata            & LAMOST \\
TYC\,1478-39-1    & 2456757.25111 & 121.55 & \nodata           & \nodata         & \nodata            & LAMOST \\
TYC\,1507-49-1    & 2457475.34315 &   0.33 & \nodata           & \nodata         & \nodata            & LAMOST \\
TYC\,1557-1803-1  & 2457497.31875 & 421.86 & 6371.38$\pm$13.33 & 4.100$\pm$0.018 & $-$0.221$\pm$0.011 & LAMOST \\
TYC\,1719-425-1   & 2456202.13169 &   9.65 & \nodata           & \nodata         & \nodata            & LAMOST \\
TYC\,1742-1301-1  & 2456551.17560 & 458.08 & 5814.84$\pm$11.86 & 4.278$\pm$0.016 & $-$0.169$\pm$0.010 & LAMOST \\
TYC\,1742-1301-1  & 2456551.21249 & 394.91 & 5828.02$\pm$12.11 & 4.300$\pm$0.017 & $-$0.159$\pm$0.010 & LAMOST \\
TYC\,1742-1301-1  & 2457327.11410 & 516.72 & 5831.32$\pm$14.65 & 4.356$\pm$0.020 & $-$0.160$\pm$0.012 & LAMOST \\
TYC\,1749-1463-1  & 2456589.20749 & 166.29 & 6381.57$\pm$15.88 & 4.207$\pm$0.026 & $-$0.038$\pm$0.014 & LAMOST \\
TYC\,1758-2133-1  & 2456202.25293 &   9.68 & \nodata           & \nodata         & \nodata            & LAMOST \\
TYC\,1761-51-1    & 2456202.22094 & 540.92 & \nodata           & \nodata         & \nodata            & LAMOST \\
TYC\,1768-162-1   & 2456675.98709 & 268.90 & \nodata           & \nodata         & \nodata            & LAMOST \\
TYC\,1783-665-1   & 2456571.27033 & 272.20 & 5532.58$\pm$17.39 & 4.383$\pm$0.026 & $-$0.748$\pm$0.014 & LAMOST \\
TYC\,1783-665-1   & 2457356.16870 & 315.84 & 5565.33$\pm$17.42 & 4.400$\pm$0.024 & $-$0.736$\pm$0.014 & LAMOST \\
TYC\,1784-1075-1  & 2456571.27034 & 426.26 & 6491.27$\pm$10.73 & 4.152$\pm$0.015 &    0.022$\pm$0.009 & LAMOST \\
TYC\,1784-1075-1  & 2457356.16872 & 513.54 & 6518.88$\pm$ 8.56 & 4.149$\pm$0.012 & $-$0.001$\pm$0.007 & LAMOST \\
TYC\,1821-1013-1  & 2456663.00263 & 359.90 & 6618.26$\pm$20.61 & 4.141$\pm$0.029 & $-$0.319$\pm$0.017 & LAMOST \\
TYC\,1821-1013-1  & 2456967.27996 & 183.26 & 6611.93$\pm$15.11 & 4.120$\pm$0.024 & $-$0.276$\pm$0.013 & LAMOST \\
TYC\,1914-31-1    & 2456280.28387 & 146.54 & 6980.44$\pm$37.64 & 4.072$\pm$0.062 &    0.291$\pm$0.034 & LAMOST \\
TYC\,1986-2176-1  & 2457003.42890 & 173.38 & 5859.00$\pm$18.25 & 4.407$\pm$0.030 & $-$0.175$\pm$0.016 & LAMOST \\
TYC\,1986-2176-1  & 2457528.03422 & 177.26 & 5825.61$\pm$21.79 & 4.340$\pm$0.035 & $-$0.229$\pm$0.019 & LAMOST \\
TYC\,2023-752-1   & 2458138.42031 & 259.92 & 5620.74$\pm$15.63 & 4.154$\pm$0.024 & $-$0.511$\pm$0.013 & LAMOST \\
TYC\,2027-86-1    & 2456063.15284 &  99.70 & 6358.69$\pm$23.44 & 4.284$\pm$0.039 & $-$0.279$\pm$0.022 & LAMOST \\
TYC\,2027-86-1    & 2456084.07977 &  77.06 & 6304.67$\pm$25.67 & 4.249$\pm$0.042 & $-$0.321$\pm$0.024 & LAMOST \\
TYC\,2036-1214-1  & 2457085.35397 & 188.02 & 6177.14$\pm$12.47 & 4.294$\pm$0.020 & $-$1.120$\pm$0.011 & LAMOST \\
TYC\,2036-1214-1  & 2458256.23926 &  43.16 & 6150.94$\pm$69.68 & 4.188$\pm$0.115 & $-$1.193$\pm$0.067 & LAMOST \\
TYC\,2298-197-1   & 2457297.27205 & 520.90 & 6699.88$\pm$ 9.60 & 4.209$\pm$0.013 & $-$0.195$\pm$0.008 & LAMOST \\
TYC\,2336-231-1   & 2456255.18835 & 456.74 & 6922.95$\pm$ 9.72 & 4.210$\pm$0.013 & $-$0.518$\pm$0.008 & LAMOST \\
TYC\,236-1252-1   & 2455974.17818 & 161.87 & 5591.08$\pm$30.89 & 4.275$\pm$0.050 & $-$0.164$\pm$0.027 & LAMOST \\
TYC\,236-1252-1   & 2457026.29595 &  94.45 & 5587.54$\pm$25.24 & 4.243$\pm$0.042 & $-$0.197$\pm$0.024 & LAMOST \\
TYC\,2472-1279-1  & 2457358.35211 & 149.82 & 6121.30$\pm$21.22 & 4.339$\pm$0.035 &    0.152$\pm$0.019 & LAMOST \\
TYC\,2506-1107-1  & 2456769.05532 &  98.86 & 5872.37$\pm$29.96 & 4.212$\pm$0.049 & $-$0.203$\pm$0.028 & LAMOST \\
\nodata & \nodata & \nodata & \nodata & \nodata& \nodata & \nodata \\
\enddata
\tablecomments{Here we list the Name, Heliocentric Julian Dates (HJD) of the RAVE/LAMOST spectrum, S/N, and stellar parameters (the \Teff, \logg, and metallicity of the companion. For RAVE data, metallicity is the [m/H], while for LAMOST, it's the [Fe/H]). The last column flags the origin of the data, i.e. RAVE or LAMOST. ``\nodata " indicates that no parameter is available. The entire table is provided in the electronic version of the paper.}
\end{deluxetable*}
\clearpage

\begin{longrotatetable}
\begin{deluxetable*}{llrrrrlrrl} 
\setlength{\tabcolsep}{2.0pt}
\tablecaption{The high resolution spectroscopic parameters of the AFGK companions in WD+AFGK binaries\label{tab:param_high_resolution} }
\tabletypesize{\scriptsize}
\tablewidth{0pt} 
\tablehead{ 
\colhead{Name} & \colhead{\Teff} & \colhead{\logg} &  \colhead{\feh} & \colhead{M} &\colhead{R} &
\colhead{\Teff $^B$} & \colhead{R$^B$} & \colhead{$A_V^B$} & \colhead{\Teff $^\mathrm{WD}$} \\
\colhead{} & \colhead{(K)} & \colhead{(dex)}  & \colhead{(dex)} & \colhead{(M$_{\sun}$)}  & \colhead{(R$_{\sun}$)} &
\colhead{(K)} & \colhead{(R$_{\sun}$)} & \colhead{(mag)} & \colhead{(K)} \\
}  
\startdata
TYC\,1006-4-1 & 4625.50$\pm$61.37 & 1.76$\pm$0.25 & $-$0.29$\pm$0.05 & 1.816$\pm$0.227 & 19.369$\pm$0.935 & 4814.62$_{-14.27} ^{+14.15}$ & 19.0436$_{-0.4094} ^{+0.4298}$ & 0.1603$_{-0.0237} ^{+0.0236}$ & 13772.61$_{-593.66} ^{+703.98}$ \\
TYC\,1223-498-1 & 5519.67$\pm$109.43 & 3.27$\pm$0.22 & $-$0.06$\pm$0.07 & 1.919$\pm$0.057 & 5.125$\pm$0.271 & 5031.45$_{-33.86} ^{+38.37}$ & 6.9760$_{-0.1120} ^{+0.1158}$ & 0.0842$_{-0.0322} ^{+0.0326}$ & 12545.90$_{-116.82} ^{+132.42}$ \\
TYC\,1385-562-1 & 6249.07$\pm$91.65 & 4.67$\pm$0.12 & 0.02$\pm$0.03 & 1.082$\pm$0.012 & 0.965$\pm$0.018 & 5855.99$_{-3.29} ^{+3.25}$ & 0.9805$_{-0.0012} ^{+0.0013}$ & 0.0011$_{-0.0008} ^{+0.0018}$ & 24959.86$_{-1384.18} ^{+1219.91}$ \\
TYC\,1394-1008-1 & 5866.53$\pm$74.29 & 4.33$\pm$0.10 & $-$0.05$\pm$0.04 & 1.006$\pm$0.038 & 1.072$\pm$0.035 & 5744.94$_{-6.34} ^{+6.83}$ & 1.1744$_{-0.0077} ^{+0.0077}$ & 0.0042$_{-0.0031} ^{+0.0065}$ & 24147.18$_{-1751.50} ^{+1770.95}$ \\
TYC\,1451-111-1 & 6468.99$\pm$177.41 & 4.27$\pm$0.31 & $-$0.09$\pm$0.11 & 1.136$\pm$0.049 & 1.142$\pm$0.041 & 6299.14$_{-19.30} ^{+15.69}$ & 1.1974$_{-0.0084} ^{+0.0085}$ & 0.0338$_{-0.0167} ^{+0.0123}$ & 15548.81$_{-891.00} ^{+975.14}$ \\
TYC\,1506-1141-1 & 4621.81$\pm$68.75 & 1.82$\pm$0.23 & $-$0.30$\pm$0.05 & 1.414$\pm$0.205 & 14.287$\pm$0.699 & 4875.68$_{-4.19} ^{+5.11}$ & 12.9464$_{-0.1192} ^{+0.1234}$ & 0.0077$_{-0.0057} ^{+0.0108}$ & 45643.73$_{-4213.62} ^{+4487.39}$ \\
TYC\,169-1942-1 & 5128.14$\pm$88.94 & 3.18$\pm$0.20 & $-$0.29$\pm$0.07 & 1.496$\pm$0.164 & 5.815$\pm$0.336 & 5050.26$_{-15.75} ^{+19.38}$ & 5.9781$_{-0.1683} ^{+0.1739}$ & 0.0182$_{-0.0128} ^{+0.0212}$ & 29638.87$_{-2315.95} ^{+2496.00}$ \\
TYC\,1707-426-1 & 5445.06$\pm$111.04 & 3.88$\pm$0.15 & $-$0.06$\pm$0.07 & 1.387$\pm$0.034 & 2.703$\pm$0.156 & 5465.98$_{-7.06} ^{+10.09}$ & 2.7895$_{-0.0211} ^{+0.0207}$ & 0.0148$_{-0.0104} ^{+0.0165}$ & 26724.66$_{-1987.48} ^{+2142.78}$ \\
TYC\,1742-1301-1 & 5781.96$\pm$128.93 & 4.22$\pm$0.16 & $-$0.24$\pm$0.08 & 0.929$\pm$0.045 & 1.015$\pm$0.036 & 5710.58$_{-10.63} ^{+14.57}$ & 1.1277$_{-0.0052} ^{+0.0053}$ & 0.0156$_{-0.0110} ^{+0.0169}$ & 21191.01$_{-1718.14} ^{+1770.96}$ \\
TYC\,1911-715-1 & 5088.65$\pm$91.69 & 3.12$\pm$0.13 & $-$0.37$\pm$0.07 & 1.209$\pm$0.126 & 4.674$\pm$0.202 & 5105.53$_{-6.15} ^{+6.87}$ & 4.5763$_{-0.0713} ^{+0.0736}$ & 0.0067$_{-0.0049} ^{+0.0101}$ & 26123.30$_{-2003.80} ^{+2120.45}$ \\
TYC\,2471-204-1 & 6040.18$\pm$189.32 & 4.12$\pm$0.28 & 0.14$\pm$0.11 & 1.162$\pm$0.059 & 1.288$\pm$0.076 & 5793.54$_{-22.89} ^{+24.01}$ & 1.4242$_{-0.0127} ^{+0.0127}$ & 0.0330$_{-0.0213} ^{+0.0236}$ & 13383.08$_{-504.33} ^{+703.20}$ \\
TYC\,2488-308-1 & 5263.17$\pm$71.98 & 3.62$\pm$0.13 & $-$0.09$\pm$0.06 & 1.262$\pm$0.035 & 2.485$\pm$0.106 & 5290.02$_{-9.49} ^{+10.86}$ & 2.5593$_{-0.0337} ^{+0.0337}$ & 0.0084$_{-0.0061} ^{+0.0125}$ & 22355.87$_{-1668.80} ^{+1716.63}$ \\
TYC\,26-39-1 & 4669.59$\pm$66.42 & 2.21$\pm$0.24 & $-$0.11$\pm$0.05 & 1.399$\pm$0.199 & 11.787$\pm$0.640 & 4694.15$_{-8.54} ^{+9.24}$ & 12.3935$_{-0.2037} ^{+0.2032}$ & 0.0256$_{-0.0164} ^{+0.0198}$ & 14971.66$_{-931.88} ^{+1037.23}$ \\
TYC\,2719-866-1 & 5002.65$\pm$84.29 & 2.62$\pm$0.11 & $-$0.20$\pm$0.07 & 1.786$\pm$0.313 & 9.789$\pm$0.541 & 4899.69$_{-10.22} ^{+10.41}$ & 12.9727$_{-0.1771} ^{+0.1823}$ & 0.1779$_{-0.0239} ^{+0.0221}$ & 14576.59$_{-759.24} ^{+891.67}$ \\
TYC\,278-239-1 & 5025.76$\pm$61.00 & 2.55$\pm$0.05 & $-$0.16$\pm$0.05 & 1.903$\pm$0.040 & 7.914$\pm$0.113 & 4953.34$_{-3.13} ^{+3.63}$ & 8.5409$_{-0.1020} ^{+0.1011}$ & 0.0043$_{-0.0032} ^{+0.0069}$ & 27815.45$_{-2045.45} ^{+2293.53}$ \\
TYC\,3029-161-1 & 5750.19$\pm$122.95 & 3.65$\pm$0.21 & $-$0.30$\pm$0.08 & 1.224$\pm$0.039 & 2.352$\pm$0.120 & 5688.84$_{-3.62} ^{+4.37}$ & 2.4765$_{-0.0139} ^{+0.0139}$ & 0.0029$_{-0.0022} ^{+0.0048}$ & 15611.92$_{-1023.82} ^{+1147.50}$ \\
TYC\,3067-471-1 & 5035.63$\pm$86.97 & 3.31$\pm$0.19 & $-$0.19$\pm$0.06 & 1.100$\pm$0.085 & 2.798$\pm$0.119 & 5124.99$_{-7.02} ^{+7.63}$ & 2.7150$_{-0.0204} ^{+0.0199}$ & 0.0116$_{-0.0082} ^{+0.0112}$ & 21808.58$_{-1523.54} ^{+1676.78}$ \\
TYC\,3080-957-1 & 5011.12$\pm$112.08 & 2.60$\pm$0.11 & $-$0.16$\pm$0.09 & 1.826$\pm$0.386 & 9.139$\pm$0.536 & 5108.10$_{-21.68} ^{+17.45}$ & 8.9227$_{-0.0604} ^{+0.0639}$ & 0.0434$_{-0.0221} ^{+0.0154}$ & 10084.66$_{-119.18} ^{+139.80}$ \\
TYC\,3094-108-1 & 5025.96$\pm$83.34 & 2.29$\pm$0.10 & $-$0.25$\pm$0.06 & 1.896$\pm$0.259 & 8.842$\pm$0.339 & 5052.48$_{-22.95} ^{+20.85}$ & 9.2675$_{-0.0949} ^{+0.0922}$ & 0.0402$_{-0.0212} ^{+0.0170}$ & 12042.18$_{-81.67} ^{+86.57}$ \\
TYC\,3097-697-1 & 4844.55$\pm$59.32 & 2.81$\pm$0.10 & $-$0.48$\pm$0.05 & 0.935$\pm$0.028 & 3.709$\pm$0.085 & 4877.60$_{-6.72} ^{+6.94}$ & 4.0443$_{-0.0226} ^{+0.0234}$ & 0.0054$_{-0.0041} ^{+0.0082}$ & 28991.00$_{-1932.04} ^{+2086.19}$ \\
TYC\,3251-1514-1 & 4832.12$\pm$38.32 & 2.55$\pm$0.13 & $-$0.13$\pm$0.05 & 1.650$\pm$0.189 & 11.230$\pm$0.374 & 4732.67$_{-15.15} ^{+14.32}$ & 12.7203$_{-0.2605} ^{+0.2665}$ & 0.0313$_{-0.0195} ^{+0.0239}$ & 13241.68$_{-483.93} ^{+618.90}$ \\
TYC\,3453-106-1 & 5294.10$\pm$72.33 & 3.13$\pm$0.08 & $-$0.21$\pm$0.07 & 1.419$\pm$0.051 & 3.341$\pm$0.159 & 5661.83$_{-7.78} ^{+8.41}$ & 2.8238$_{-0.0309} ^{+0.0305}$ & 0.0036$_{-0.0027} ^{+0.0059}$ & 30346.15$_{-2344.61} ^{+2607.33}$ \\
TYC\,3457-852-1 & 5179.28$\pm$63.73 & 3.32$\pm$0.14 & $-$0.20$\pm$0.04 & 1.277$\pm$0.050 & 2.904$\pm$0.113 & 5173.64$_{-6.92} ^{+7.50}$ & 2.9022$_{-0.0231} ^{+0.0227}$ & 0.0072$_{-0.0051} ^{+0.0103}$ & 23960.14$_{-1767.68} ^{+1756.03}$ \\
TYC\,3464-912-1 & 6025.19$\pm$70.62 & 4.74$\pm$0.04 & $-$0.26$\pm$0.06 & 0.969$\pm$0.038 & 1.110$\pm$0.032 & 6029.25$_{-6.79} ^{+7.09}$ & 0.9970$_{-0.0002} ^{+0.0002}$ & 0.0017$_{-0.0013} ^{+0.0029}$ & 20827.18$_{-1443.20} ^{+1519.67}$ \\
TYC\,3640-1105-1 & 6338.37$\pm$230.87 & 4.19$\pm$0.21 & $-$0.09$\pm$ 0.13 & 1.284$\pm$0.074 & 1.692$\pm$0.125 & 5969.77$_{-5.74} ^{+6.92}$ & 1.9378$_{-0.0126} ^{+0.0129}$ & 0.0047$_{-0.0035} ^{+0.0071}$ & 42266.62$_{-3888.82} ^{+3803.86}$ \\
TYC\,368-1591-1 & 4564.74$\pm$75.95 & 1.29$\pm$0.28 & $-$0.45$\pm$0.07 & 1.294$\pm$0.242 & 19.543$\pm$0.891 & 4704.78$_{-7.72} ^{+7.68}$ & 20.6288$_{-0.2459} ^{+0.2484}$ & 0.1908$_{-0.0253} ^{+0.0235}$ & 24547.35$_{-1881.66} ^{+1963.13}$ \\
TYC\,3807-183-1 & 5452.71$\pm$71.24 & 3.04$\pm$0.08 & $-$0.59$\pm$0.08 & 1.863$\pm$0.057 & 6.274$\pm$0.287 & 5441.60$_{-3.63} ^{+4.76}$ & 6.1730$_{-0.1344} ^{+0.1377}$ & 0.0043$_{-0.0032} ^{+0.0065}$ & 76159.38$_{-6904.40} ^{+6939.21}$ \\
TYC\,3808-1388-1 & 4784.63$\pm$76.06 & 2.72$\pm$0.20 & $-$0.08$\pm$0.06 & 1.394$\pm$0.146 & 8.110$\pm$0.415 & 4844.79$_{-8.90} ^{+8.44}$ & 7.8196$_{-0.1324} ^{+0.1341}$ & 0.0608$_{-0.0226} ^{+0.0223}$ & 24980.53$_{-1825.22} ^{+1925.63}$ \\
TYC\,3816-790-1 & 4602.23$\pm$65.74 & 1.76$\pm$0.25 & $-$0.25$\pm$0.04 & 1.192$\pm$0.151 & 11.412$\pm$0.758 & 4716.33$_{-6.51} ^{+6.51}$ & 11.4287$_{-0.1447} ^{+0.1468}$ & 0.0072$_{-0.0054} ^{+0.0098}$ & 22035.62$_{-1615.52} ^{+1638.69}$ \\
TYC\,3829-462-1 & 5872.74$\pm$162.5 & 4.52$\pm$0.16 & $-$0.33$\pm$0.10 & 0.884$\pm$0.041 & 0.855$\pm$0.027 & 5617.50$_{-6.24} ^{+7.38}$ & 1.0336$_{-0.0018} ^{+0.0018}$ & 0.0114$_{-0.0079} ^{+0.0098}$ & 21892.30$_{-1713.74} ^{+1747.86}$ \\
TYC\,3841-492-1 & 4871.77$\pm$50.17 & 2.75$\pm$0.21 & $-$0.35$\pm$0.04 & 1.333$\pm$0.14 & 10.827$\pm$0.255 & 4841.72$_{-3.23} ^{+3.48}$ & 11.4237$_{-0.0694} ^{+0.0704}$ & 0.0045$_{-0.0033} ^{+0.0066}$ & 21404.70$_{-1653.05} ^{+1706.91}$ \\
TYC\,3868-840-1 & 4571.12$\pm$74.05 & 1.81$\pm$0.33 & $-$0.38$\pm$0.08 & 1.084$\pm$0.115 & 11.424$\pm$0.585 & 4690.68$_{-5.07} ^{+5.64}$ & 11.3052$_{-0.1171} ^{+0.1159}$ & 0.0094$_{-0.0069} ^{+0.0121}$ & 23682.25$_{-1797.60} ^{+1915.22}$ \\
TYC\,3881-159-1 & 6044.66$\pm$182.18 & 4.14$\pm$0.28 & $-$0.11$\pm$0.14 & 1.011$\pm$0.061 & 1.038$\pm$0.045 & 6080.76$_{-8.34} ^{+8.73}$ & 0.9597$_{-0.0022} ^{+0.0022}$ & 0.0030$_{-0.0022} ^{+0.0049}$ & 26083.32$_{-1533.22} ^{+1629.60}$ \\
TYC\,3883-1104-1 & 5041.81$\pm$87.45 & 3.04$\pm$0.15 & $-$0.18$\pm$0.07 & 1.289$\pm$0.125 & 4.647$\pm$0.208 & 5075.79$_{-20.61} ^{+20.46}$ & 4.3722$_{-0.0659} ^{+0.0647}$ & 0.0186$_{-0.0133} ^{+0.0197}$ & 26823.48$_{-1922.99} ^{+2115.15}$ \\
TYC\,405-806-1 & 4779.99$\pm$103.41 & 2.22$\pm$0.32 & $-$0.23$\pm$0.10 & 1.097$\pm$0.116 & 5.365$\pm$0.249 & 4596.19$_{-22.96} ^{+22.20}$ & 6.9271$_{-0.0996} ^{+0.0998}$ & 0.1980$_{-0.0298} ^{+0.0288}$ & 17081.07$_{-1179.26} ^{+1209.56}$ \\
TYC\,4102-715-1 & 5137.58$\pm$90.20 & 3.31$\pm$0.20 & $-$0.42$\pm$0.07 & 1.082$\pm$0.101 & 3.120$\pm$0.132 & 4978.02$_{-12.94} ^{+12.63}$ & 3.6467$_{-0.0420} ^{+0.0414}$ & 0.1519$_{-0.0261} ^{+0.0257}$ & 27642.26$_{-2219.99} ^{+2397.45}$ \\
TYC\,4219-2017-1 & 4933.69$\pm$65.93 & 2.63$\pm$0.13 & $-$0.28$\pm$0.04 & 1.114$\pm$0.103 & 4.503$\pm$0.160 & 5054.83$_{-13.58} ^{+17.75}$ & 4.2871$_{-0.0379} ^{+0.0377}$ & 0.0164$_{-0.0118} ^{+0.0197}$ & 27689.55$_{-1995.40} ^{+2104.84}$ \\
TYC\,4220-740-1 & 4663.71$\pm$74.65 & 2.62$\pm$0.21 & $-$0.03$\pm$0.09 & 1.108$\pm$0.096 & 5.585$\pm$0.220 & 4801.32$_{-14.73} ^{+13.99}$ & 5.4827$_{-0.0455} ^{+0.0455}$ & 0.0661$_{-0.0235} ^{+0.0233}$ & 19012.61$_{-1354.81} ^{+1399.74}$ \\
TYC\,4352-264-1 & 4577.29$\pm$58.63 & 1.85$\pm$0.20 & $-$0.32$\pm$0.05 & 1.562$\pm$0.177 & 17.183$\pm$0.703 & 5199.71$_{-34.61} ^{+34.46}$ & 13.4281$_{-0.1347} ^{+0.1395}$ & 0.1919$_{-0.0319} ^{+0.0323}$ & 9843.75$_{-46.24} ^{+46.53}$ \\
TYC\,4385-1146-1 & 6207.12$\pm$49.36 & 4.44$\pm$0.09 & $-$0.52$\pm$0.08 & 0.945$\pm$0.039 & 0.954$\pm$0.018 & 6009.88$_{-4.27} ^{+5.95}$ & 0.9565$_{-0.0023} ^{+0.0023}$ & 0.0035$_{-0.0026} ^{+0.0057}$ & 28351.92$_{-1642.59} ^{+1816.02}$ \\
TYC\,4442-1466-1 & 4823.69$\pm$71.36 & 1.98$\pm$0.12 & $-$0.49$\pm$0.06 & 1.511$\pm$0.233 & 13.627$\pm$0.649 & 4921.83$_{-5.67} ^{+7.18}$ & 13.4331$_{-0.1111} ^{+0.1143}$ & 0.0235$_{-0.0153} ^{+0.0195}$ & 35113.69$_{-3109.63} ^{+3494.33}$ \\
TYC\,4559-684-1 & 4727.17$\pm$86.43 & 2.20$\pm$0.25 & $-$0.20$\pm$0.09 & 1.416$\pm$0.219 & 11.601$\pm$0.677 & 4777.10$_{-4.51} ^{+5.23}$ & 12.0509$_{-0.1175} ^{+0.1170}$ & 0.0091$_{-0.0066} ^{+0.0124}$ & 28701.19$_{-2284.57} ^{+2466.28}$ \\
TYC\,4564-627-1 & 6425.64$\pm$130.34 & 4.93$\pm$0.05 & $-$0.30$\pm$0.10 & 1.097$\pm$0.054 & 1.175$\pm$0.045 & 6260.69$_{-22.35} ^{+17.54}$ & 1.2987$_{-0.0066} ^{+0.0065}$ & 0.1174$_{-0.0220} ^{+0.0169}$ & 82073.84$_{-7220.90} ^{+5350.56}$ \\
TYC\,4574-366-1 & 4866.91$\pm$79.95 & 2.59$\pm$0.14 & $-$0.12$\pm$0.06 & 1.302$\pm$0.130 & 6.240$\pm$0.259 & 4847.98$_{-6.43} ^{+8.01}$ & 6.4107$_{-0.0312} ^{+0.0340}$ & 0.0219$_{-0.0145} ^{+0.0196}$ & 19247.88$_{-1480.54} ^{+1556.12}$ \\
TYC\,4615-1151-1 & 4678.50$\pm$75.22 & 2.15$\pm$0.23 & $-$0.31$\pm$0.06 & 1.259$\pm$0.173 & 11.346$\pm$0.688 & 4711.35$_{-7.93} ^{+8.10}$ & 13.5254$_{-0.1522} ^{+0.1545}$ & 0.1445$_{-0.0210} ^{+0.0216}$ & 26950.68$_{-2095.20} ^{+2192.92}$ \\
TYC\,4649-3689-1 & 5086.06$\pm$77.21 & 2.81$\pm$0.26 & 0.08$\pm$0.07 & 2.814$\pm$0.093 & 14.284$\pm$0.478 & 5014.5$_{-11.25} ^{+17.96}$ & 14.0749$_{-0.1647} ^{+0.1692}$ & 0.0195$_{-0.0130} ^{+0.0212}$ & 21503.37$_{-1586.80} ^{+1718.39}$ \\
TYC\,4665-621-1 & 4735.21$\pm$64.45 & 2.64$\pm$0.08 & $-$0.02$\pm$0.05 & 1.361$\pm$0.121 & 7.646$\pm$0.330 & 4762.97$_{-7.63} ^{+7.74}$ & 7.5859$_{-0.1232} ^{+0.1250}$ & 0.0119$_{-0.0088} ^{+0.0153}$ & 18703.70$_{-1228.61} ^{+1315.26}$ \\
TYC\,4681-1527-1 & 5432.54$\pm$122.25 & 3.14$\pm$0.18 & $-$0.24$\pm$0.09 & 1.364$\pm$0.055 & 3.007$\pm$0.246 & 5669.19$_{-4.58} ^{+7.40}$ & 2.8611$_{-0.0839} ^{+0.0850}$ & 0.0066$_{-0.0049} ^{+0.0101}$ & 28520.66$_{-2313.95} ^{+2389.20}$ \\
TYC\,4685-1113-1 & 4748.91$\pm$57.54 & 2.49$\pm$0.14 & $-$0.19$\pm$0.06 & 1.897$\pm$0.267 & 14.203$\pm$0.945 & 4762.70$_{-8.85} ^{+10.03}$ & 14.4456$_{-0.5439} ^{+0.6254}$ & 0.0164$_{-0.0116} ^{+0.0189}$ & 40773.74$_{-3563.79} ^{+3702.39}$ \\
TYC\,4698-895-1 & 5031.75$\pm$93.31 & 3.13$\pm$0.15 & 0.10$\pm$0.07 & 1.285$\pm$0.094 & 3.127$\pm$0.170 & 4884.34$_{-20.98} ^{+22.37}$ & 3.3165$_{-0.0477} ^{+0.0489}$ & 0.0254$_{-0.0180} ^{+0.0244}$ & 14663.98$_{-624.36} ^{+772.14}$ \\
TYC\,4717-255-1 & 4669.64$\pm$64.60 & 2.27$\pm$0.21 & $-$0.14$\pm$0.05 & 2.038$\pm$0.206 & 17.590$\pm$0.783 & 4880.21$_{-27.20} ^{+23.15}$ & 17.1540$_{-0.3098} ^{+0.3213}$ & 0.2326$_{-0.0320} ^{+0.0289}$ & 11066.93$_{-149.24} ^{+169.69}$ \\
TYC\,4831-473-1 & 5875.75$\pm$91.00 & 3.52$\pm$0.14 & $-$0.23$\pm$0.08 & 0.940$\pm$0.044 & 1.048$\pm$0.035 & 5869.21$_{-21.86} ^{+23.50}$ & 1.1074$_{-0.0055} ^{+0.0054}$ & 0.0551$_{-0.0263} ^{+0.0281}$ & 29737.52$_{-2483.18} ^{+2523.61}$ \\
TYC\,5-436-1 & 4781.09$\pm$71.07 & 2.47$\pm$0.09 & $-$0.28$\pm$0.06 & 1.697$\pm$0.263 & 13.607$\pm$0.732 &  4811.54$_{-8.04} ^{+6.67}$ & 14.2855$_{-0.3002} ^{+0.3045}$ & 0.0608$_{-0.0178} ^{+0.0120}$ & 49015.12$_{-4754.36} ^{+4668.07}$ \\
TYC\,856-918-1 & 4855.40$\pm$70.04 & 2.37$\pm$0.09 & $-$0.21$\pm$0.05 & 1.304$\pm$0.243 & 10.522$\pm$0.422 & 4872.50$_{-11.48} ^{+10.91}$ & 11.1999$_{-0.2749} ^{+0.2670}$ & 0.0732$_{-0.0247} ^{+0.0240}$ & 44185.79$_{-4261.46} ^{+4312.38}$ \\
TYC\,877-681-1 & 5034.89$\pm$89.67 & 2.63$\pm$0.18 & $-$0.19$\pm$0.07 & 1.928$\pm$0.071 & 8.088$\pm$0.230 & 5045.09$_{-18.02} ^{+20.76}$ & 8.1619$_{-0.1412} ^{+0.1444}$ & 0.0187$_{-0.0133} ^{+0.0218}$ & 11744.89$_{-143.00} ^{+152.26}$ \\
\enddata
\tablecomments{Here we  list the  stellar  parameters (\Teff, \logg, [Fe/H]), mass and radius determined from  high resolution spectra, the \Teff$^B$, R$^B$, $A_V^B$ of AFGK companions and $T_\mathrm{eff}^\mathrm{WD}$ of WD estimated from binary SED fitting solution.}
\end{deluxetable*}
\end{longrotatetable}


\bibliographystyle{aasjournal}
\bibliography{ms}{}

\end{document}